\pdfoutput=1

\documentclass[preprint,journal]{vgtc}            %

\onlineid{6505}

\ieeedoi{10.1109/TVCG.2025.3567132}

\vgtccategory{Research}

\vgtcpapertype{Theory/Model}

\renewcommand{\manuscriptnotetxt}{© 2025 IEEE.
This is the author's version of the article that has been accepted in IEEE TVCG.
Personal use of this material is permitted.
Permission from IEEE must be obtained for all other uses, in any current or future media, including reprinting/republishing this material for advertising or promotional purposes, creating new collective works, for resale or redistribution to servers or lists, or reuse of any copyrighted component of this work in other works.
}

\usepackage{multirow}

\newcommand{\inlinecode}[1]{\texttt{\detokenize{#1}}}
\newcommand{\inlinecodesmall}[1]{{\fontsize{8}{12}\texttt{\detokenize{#1}}}}

\setlist{noitemsep,parsep=0pt,partopsep=0pt}

\usepackage[group-separator={$,$}]{siunitx}

\newcommand{\systemName}{VisTaxa\xspace}

\newcommand{\nOldVis}{13511\xspace}
\newcommand{\nOldVisShort}{13K\xspace}
\newcommand{\nOldVisLabeled}{400\xspace}
\newcommand{\nOldVisPredicted}{13111\xspace}
\newcommand{\nOldVisPublishYearUnknown}{101\xspace}

\newcommand{\nGlossaryPaper}{10\xspace}

\newcommand{\coderIndexWeiliZheng}{C1\xspace}
\newcommand{\coderIndexXinyuChen}{C2\xspace}
\newcommand{\coderIndexYuZhang}{C3\xspace}

\newcommand{\nTaxaFinalAllVis}{\num{51}\xspace}
\newcommand{\nTaxaFinalLeafVis}{\num{45}\xspace}
\newcommand{\nTaxaFinalFirstLevelVis}{\num{20}\xspace}

\newcommand{\taxon}{\textsl}

\newcommand{\leafTaxon}{\textbf}

\newcommand{\scheme}{\texttt}

\title{VisTaxa: Developing a Taxonomy of Historical Visualizations}

\author{
    \authororcid{Yu~Zhang}{0000-0002-9035-0463},
    \authororcid{Xinyue~Chen}{0009-0006-0700-4575},
    \authororcid{Weili~Zheng}{0009-0000-4209-9094},
    \authororcid{Yuhan~Guo}{0009-0004-3857-7486},\\
    \authororcid{Guozheng~Li}{0000-0001-6663-6712},
    \authororcid{Siming~Chen}{0000-0002-2690-3588},
    and \authororcid{Xiaoru~Yuan}{0000-0002-7233-980X}
}

\authorfooter{
    \item
    Yu~Zhang is with the Department of Computer Science, University of Oxford.
    E-mail: yuzhang94@outlook.com.

    \item
    Xinyue~Chen, Yuhan~Guo, and Xiaoru~Yuan (corresponding) are with the National Key Laboratory of General Artificial Intelligence, and School of Intelligence Science and Technology, Peking University and PKU-WUHAN Institute for Artificial Intelligence.
    E-mail: cxyapril@stu.pku.edu.cn and \{yuhan.guo | xiaoru.yuan\}@pku.edu.cn.

    \item
    Weili~Zheng and Siming~Chen (corresponding) are with the School of Data Science, Fudan University.
    E-mail: \{wlzheng17 | simingchen\}@fudan.edu.cn.

    \item
    Guozheng~Li is with Beijing Institute of Technology.
    E-mail: guozheng.li@bit.edu.cn.
}

\abstract{%
  Historical visualizations are a rich resource for visualization research.
While taxonomy is commonly used to structure and understand the design space of visualizations, existing taxonomies primarily focus on contemporary visualizations and largely overlook historical visualizations.
To address this gap, we describe an empirical method for taxonomy development.
We introduce a coding protocol and the \systemName system for taxonomy labeling and comparison.
We demonstrate using our method to develop a historical visualization taxonomy by coding \nOldVisLabeled images of historical visualizations.
We analyze the coding result and reflect on the coding process.
Our work is an initial step toward a systematic investigation of the design space of historical visualizations.

}


\graphicspath{{figs/}{figures/}{pictures/}{images/}{./}} %

\usepackage{tabu}                      %
\usepackage{booktabs}                  %
\usepackage{lipsum}                    %
\usepackage{mwe}                       %

\usepackage{mathptmx}                  %

\begin{document}

\ifx\hidemain\undefined
    \maketitle

    \section{Introduction}

Historical visualizations have long been a valuable source of inspiration for contemporary visualization research, which stimulates discussions on good and bad practices of visualization design~\cite{Tufte1983Visual} and informs the development of visualization grammar~\cite{Bertin1983Semiology}.
Meanwhile, historical visualizations are voluminous, diverse, and disarrayed, making them difficult to retrieve and comprehend.
To promote a more effective use of historical visualizations, efforts have been made to gather them systematically, such as the early Milestones project~\cite{Friendly2001Milestones} and the recent OldVisOnline dataset~\cite{Zhang2024OldVisOnline}.

This work aims to study the design space of historical visualizations as part of the ongoing effort to provide a more structured understanding of them.
We resort to taxonomy to systematically describe what historical visualizations comprise.
The emergence of massive online digital libraries~\cite{RumseyDavid} and dedicated historical visualization collections~\cite{Zhang2024OldVisOnline} presents a particular opportunity to empirically build a taxonomy of historical visualization using such data.

Visualization taxonomy has long been studied for structuring the design space~\cite{Shneiderman1996Eyes,Card1997Structure}.
The taxonomy itself serves as a structured overview of the design space and the disciplinary landscape.
It provides a way to group images and supports statistical analysis, such as the temporal distribution of each visual design.
Taxonomy labels can be used as keywords for retrieval to support better utilization of the visualization resource.
Meanwhile, existing taxonomies primarily target contemporary visualizations, which may not align with the distribution of visual designs in historical visualizations.
Thus, these taxonomies may not be the best fit for the usage above for historical visualizations compared to a taxonomy dedicated to historical visualizations.

As a step toward a comprehensive taxonomy of historical visualizations, this work describes an empirical method to develop the taxonomy through qualitative coding (\cref{sec:coding-method}).
We systematize and operationalize the coding process to reinforce analytic transparency~\cite{Moravcsik2019Transparency}.
Our coding protocol borrows techniques from grounded theory~\cite{Glaser1967Discovery}, such as constant comparison, memo writing, and theoretical saturation.
During coding, the coders iteratively refine the taxonomy tree and assign taxa to the visualization images.
To support this process, we have developed a system named \systemName that integrates machine assistance for image taxonomy labeling and comparison (\cref{sec:system}).
Our protocol and \systemName system are designed to accommodate corpus update and batching, allowing the taxonomy to be incrementally updated.

We used our protocol and system to develop a preliminary taxonomy of historical visualizations in the OldVisOnline~\cite{Zhang2024OldVisOnline} dataset (\cref{sec:result-and-analysis}).
Three of the authors coded \nOldVisLabeled historical visualization images and obtained a taxonomy with \nTaxaFinalAllVis visualization taxa.
Using the taxonomy labels, we predict the labels of the remaining \nOldVisShort images in OldVisOnline, which allows us to obtain a coarse overview of the temporal distribution of visual designs throughout history.
We also compare our taxonomy with prior visual representation classification schemes.
Furthermore, reflecting on the coding process (\cref{sec:reflection}),
we discuss observations on the iterative development of the taxonomy and the consensus and dissensus among coders.

Our work is a preliminary step toward systematically investigating the following research question: \textit{What visual designs have been used in historical visualizations?}
This work has the following contributions:

\begin{itemize}[leftmargin=*]
    \item We describe a method for developing a taxonomy of visualization images.
          It involves a coding protocol and the \systemName system for taxonomy labeling and comparison.

    \item We use our method to develop a taxonomy for historical visualizations.
          We discuss findings and reflect on the coding process.
\end{itemize}

The taxonomy labels are available at \url{https://github.com/oldvis/image-taxonomy}.
\systemName's source code is available at \url{https://github.com/oldvis/image-taxonomy-labeler}.

    \section{Related Work}

This section reviews related work on visualization representation taxonomy.
\Cref{sec:characterizing-schemes} characterizes existing classification schemes of visual representations to position our taxonomy and identify the most relevant works.
\Cref{sec:why-taxonomy} reviews scenarios that motivate taxonomies.
\Cref{sec:how-to-develop-taxonomy} discusses methods for developing a taxonomy.

\subsection{Visual Representation Classification Schemes}
\label{sec:characterizing-schemes}

\Cref{tab:taxonomies} categorizes existing classification schemes\footnote{
    The literature has used ``taxonomy'' to refer to related while different schemes for organizing entities due to the different adopted definitions~\cite{Smith2002Typologies,Jacob2004Classification}.
    We use ``classification scheme'' to refer to these notions of ``taxonomy'' and related concepts (e.g., typology, classification, and categorization) in general.
    When referring to a scheme, we use the term in the original literature.
} and positions our taxonomy (referred to as \scheme{Ours}) in the following aspects.

\begin{itemize}[leftmargin=*]
    \item \textbf{Theme:}
          Existing schemes focus on various themes.
          We categorize the themes by \textit{scope} and \textit{subject}.

          \begin{itemize}[leftmargin=*]
              \item \textbf{Scope:}
                    The scope of entities to be included in a scheme differs.
                    The scheme may focus on visualizations of specific types (e.g., timeline~\cite{Brehmer2017Timelines}, tree~\cite{Schulz2011Treevis.net}, text~\cite{Kucher2015Text}) or visualizations in general~\cite{Lohse1994Classification,Chen2022Not}.
                    \scheme{Ours} focuses on the latter.

              \item \textbf{Subject: technique or instance.}
                    A subtle difference is whether the scheme categorizes techniques or concrete instances (e.g., visualization images).
                    \scheme{Ours} focuses on categorizing instances.
          \end{itemize}

    \item \textbf{Organization structure:}
          The scheme may organize entities with different structures.
          We categorize these structures by \textit{dimensionality}, \textit{depth}, and \textit{relation}.

          \begin{itemize}[leftmargin=*]
              \item \textbf{Dimensionality: 1D or faceted.}
                    A scheme can be one-dimensional (1D) or faceted, with multiple dimensions.
                    Examples of 1D schemes include Lohse et al.'s classification of visual representations~\cite{Lohse1994Classification} and Chen et al.'s typology of IEEE VIS images~\cite{Chen2022Not}.
                    Faceted schemes are commonly used to present orthogonal dimensions of a design space.
                    Examples of faceted schemes include the task by data type taxonomy~\cite{Shneiderman1996Eyes} and the timeline design space with the facets of representation, scale, and layout~\cite{Brehmer2017Timelines}.
                    \scheme{Ours} focuses on 1D taxonomy.

              \item \textbf{Depth: single-level or multi-level.}
                    A 1D scheme (or a dimension of a faceted scheme) can be single-level or multi-level.
                    We exclude the root node when counting the depth.
                    We also exclude the leaf node when the categorized subject is instance, in which case the leaf nodes are instances.
                    \scheme{Ours} focuses on multi-level taxonomy.

              \item \textbf{Relation: is-a or others.}
                    A multi-level taxonomy typically represents a hierarchy with an ``is-a'' relation, where the concept associated with a child node is a subset of the concept associated with its parent node~\cite{Brachman1983What}.
                    Other schemes may represent different relations.
                    For example, a child node and its parent node may denote values of different facets~\cite{Hograefer2020State}.
                    \scheme{Ours} represents is-a relation.
          \end{itemize}

    \item \textbf{Methodological approach:}
          Different methodological approaches have been used in the literature to develop a scheme.
          We categorize these approaches by \textit{process} and \textit{criteria}.

          \begin{itemize}[leftmargin=*]
              \item \textbf{Process: empirical, conceptual, or mixed.}
                    A scheme can be developed empirically, which may involve a process of data collection and qualitative analysis of observations.
                    The scheme can also be developed conceptually, presented as abstract dimensions (e.g., Bertin's taxonomy of visual variables~\cite{Bertin1983Semiology}).
                    The process can also be mixed, where empirical observations inform and validate the design of abstract dimensions (e.g., Brehmer's timeline design space~\cite{Brehmer2017Timelines}).
                    The literature does not always clarify the taxonomy development process. Our categorization is based on our best guesses.
                    Note that taxonomy is sometimes defined to be empirical, while the conceptual counterpart is referred to as typology~\cite{Smith2002Typologies}.
                    \scheme{Ours} is developed empirically using a dataset of visualizations.

              \item \textbf{Criteria: structural, functional, or mixed.}
                    The criteria for developing a taxonomy can be structural or functional, as summarized by Rankin~\cite{Rankin1990Taxonomy}.
                    Structural taxonomy focuses on the form.
                    Functional taxonomy focuses on the use.
                    The structural and functional criteria may be used jointly.
                    For example, Lo et al.'s taxonomy categorizes issues in misleading visualizations into stages of the visualization process, which includes issues in processing the input data (i.e., a functional aspect) and the visualization design (i.e., a structural aspect)~\cite{Lo2022Misinformed}.
                    \scheme{Ours} is a structural taxonomy.

          \end{itemize}
\end{itemize}

Among the above aspects, \textit{criteria} is specific to schemes for visual representation, while the others apply to schemes in general.

With the characterization of schemes summarized in \cref{tab:taxonomies}, we see the related work~\cite{Lohse1994Classification,Friendly2001Milestones,Chen2022Not,Chen2024Image} categorizing visualization instances closest to \scheme{Ours}.
\Cref{sec:comparison-with-existing-schemes} compares \scheme{Ours} with them.

\begin{table}[!htbp]
    \centering
    \scriptsize
    \caption{
        \textbf{Characteristics of visual representation classification schemes:}
        The characterization is based on theme, organization structure, and methodological approach (details in \cref{sec:characterizing-schemes}).
        When the \textit{depth} is single-level, there is no hierarchical \textit{relation} (denoted as ``/'').
        ``vis.'' abbreviates ``visualization''.
        ``rep.'' abbreviates ``representation''.
    }
    \setlength{\aboverulesep}{0pt}
    \setlength{\belowrulesep}{0pt}
    \label{tab:taxonomies}
    \scalebox{0.67}{
        \setlength{\tabcolsep}{0.25em}
        \begin{tabular}{c|cc|ccc|cc}
            \toprule
            \multirow{2}[0]{*}{\textbf{Scheme}}                    & \multicolumn{2}{c|}{\textbf{Theme}} & \multicolumn{3}{c|}{\textbf{Structure}} & \multicolumn{2}{c}{\textbf{Approach}}                                                                                                                                                              \\\cline{2-8}
                                                                   & \multicolumn{1}{c}{\textbf{Scope}}  & \multicolumn{1}{c|}{\textbf{Subject}}   & \multicolumn{1}{c}{\textbf{Dim}}      & \multicolumn{1}{c}{\textbf{Depth}} & \multicolumn{1}{c|}{\textbf{Relation}} & \multicolumn{1}{c}{\textbf{Process}} & \multicolumn{1}{c}{\textbf{Criteria}} \\
            \midrule
            \scheme{Macdonald-Ross1977}\cite{MacdonaldRoss1977How} & quantitative data rep.              & technique                               & 1D                                    & multi                              & others                                 & empirical                            & mixed                                 \\
            \scheme{Wehrend1990}~\cite{Wehrend1990Problem}         & scientific vis.                     & technique                               & faceted                               & single                             & /                                      & mixed                                & mixed                                 \\
            \scheme{Rankin1990}~\cite{Rankin1990Taxonomy}          & coordinate-based graph              & technique                               & faceted                               & single                             & /                                      & conceptual                           & structural                            \\
            \scheme{Lohse1994}~\cite{Lohse1994Classification}      & visual rep.                         & instance                                & 1D                                    & multi                              & is-a                                   & empirical                            & structural                            \\
            \scheme{Shneiderman1996}~\cite{Shneiderman1996Eyes}    & information vis.                    & technique                               & faceted                               & single                             & /                                      & mixed                                & mixed                                 \\
            \scheme{Fujishiro2000}~\cite{Fujishiro2000GADGET_IV}   & information vis.                    & technique                               & faceted                               & single                             & /                                      & mixed                                & mixed                                 \\
            \scheme{Friendly2001}~\cite{Friendly2001Milestones}    & vis. milestones                     & instance                                & faceted                               & multi                              & is-a                                   & empirical                            & mixed                                 \\
            \scheme{Draper2009}~\cite{Draper2009Survey}            & radial vis.                         & technique                               & 1D                                    & multi                              & is-a                                   & empirical                            & structural                            \\
            \scheme{Reimer2010}~\cite{Reimer2010Understanding}     & chorematic map                      & technique                               & 1D                                    & single                             & /                                      & conceptual                           & structural                            \\
            \scheme{Schulz2011}~\cite{Schulz2011Treevis.net}       & tree vis.                           & technique                               & faceted                               & single                             & /                                      & mixed                                & structural                            \\
            \scheme{Borkin2013}~\cite{Borkin2013What}              & vis. from web                       & instance                                & 1D                                    & multi                              & is-a                                   & empirical                            & structural                            \\
            \scheme{Kucher2015}~\cite{Kucher2015Text}              & text vis.                           & technique                               & faceted                               & multi                              & others                                 & empirical                            & mixed                                 \\
            \scheme{Brehmer2017}~\cite{Brehmer2017Timelines}       & timeline vis.                       & instance                                & faceted                               & single                             & /                                      & mixed                                & structural                            \\
            \scheme{Battle2018}~\cite{Battle2018Beagle}            & vis. from web                       & instance                                & 1D                                    & single                             & /                                      & empirical                            & structural                            \\
            \scheme{Hogräfer2020}~\cite{Hograefer2020State}        & map-like vis.                       & technique                               & 1D                                    & multi                              & others                                 & mixed                                & structural                            \\
            \scheme{Lo2022}~\cite{Lo2022Misinformed}               & misleading vis.                     & instance                                & 1D                                    & multi                              & is-a                                   & empirical                            & mixed                                 \\
            \scheme{Deng2023}~\cite{Deng2023VisImages}             & scholarly vis.                      & instance                                & 1D                                    & multi                              & is-a                                   & empirical                            & structural                            \\
            \scheme{Chen2024}~\cite{Chen2022Not,Chen2024Image}     & scholarly vis.                      & instance                                & 1D                                    & multi                              & is-a                                   & empirical                            & structural                            \\
            \textbf{\scheme{Ours}}                                 & historical vis.                     & instance                                & 1D                                    & multi                              & is-a                                   & empirical                            & structural                            \\
            \bottomrule
        \end{tabular}
    }
\end{table}

\subsection{Why a New Taxonomy}
\label{sec:why-taxonomy}

The following is an incomplete list of usage scenarios that motivate visualization taxonomies.

\textbf{Serve as a disciplinary overview:}
Taxonomies are widely used to structure the visualization design space and explore new design opportunities.
As a structured summary of existing artifacts, a taxonomy helps the audience understand the disciplinary landscape.
The structured summary may also help standardize the discipline~\cite{Lohse1994Classification,Chen2022Not}.

\textbf{Facilitate analysis:}
A taxonomy groups items and thus facilitates further analysis, such as examining trends and evolution over time.
For example, Schulz used a taxonomy to trace the evolution of radially stacked tree visualization~\cite{Schulz2011Treevis.net}.

\textbf{Support retrieval:}
A taxonomy associates visualization techniques or instances with taxa.
The taxa can serve as keywords for searching and retrieving, which promotes the use of visualization resources.
For example, Friendly's categorization allows readers to examine milestones in visualization history by content and form~\cite{Friendly2001Milestones}.
Schulz describes three specific scenarios of searching visualizations: retrieve a known design for exact information, search for similar designs of a given design, and search for a suitable technique for a given dataset or application~\cite{Schulz2011Treevis.net}.

Bearing these usage scenarios in mind, our taxonomy of historical visualization also aims to provide an overview of the design space (\cref{fig:taxonomy}) and facilitate analysis (\cref{sec:distribution-analysis}) of historical visualizations.
Our taxonomy labels may be used as keywords to retrieve historical visualizations in OldVisOnline~\cite{Zhang2024OldVisOnline}.
Compared with existing taxonomies of contemporary visualizations, a taxonomy tailored for historical visualizations may better serve these purposes.
Designs that are common in historical visualizations but uncommon nowadays may not exist as a taxon in the taxonomy of contemporary visualizations.

\subsection{How to Develop a Taxonomy}
\label{sec:how-to-develop-taxonomy}

Further to \cref{sec:characterizing-schemes}'s summary of methodological approaches, the following describes specific methods for developing taxonomies.

One of the ways to develop taxonomies is by conducting expert interviews~\cite{Hograefer2020State}.
Another way is through qualitative data analysis, which may involve examining a visualization corpus and applying qualitative coding methods~\cite{Lo2022Misinformed,Diehl2024Analysis}.
A taxonomy does not need to be built from scratch.
It may be developed deductively by applying a part or the whole of an existing classification scheme~\cite{Fujishiro2000GADGET_IV,Nusrat2016State,Hograefer2020State}.
Qualitative coding usually involves multiple coders.
Strategies such as voting~\cite{Lohse1994Classification} or discussions to reach consensus~\cite{Chen2022Not} are used to resolve conflicts between the results of different coders.

As the method for taxonomy development is usually not the focus of related work, the steps and tools used are often obscure.
We systematize and operationalize a coding method for visualization taxonomy development (\cref{sec:coding-method}) and contribute a system for taxonomy labeling and comparison (\cref{sec:system}).
Our system aligns with the thread of work on systems for qualitative coding~\cite{Drouhard2017Aeonium,Chandrasegaran2017Integrating,Rietz2021Cody}, but differs in that we focus on coding images, while prior work mainly focuses on text.

    \begin{figure*}[!htbp]
    \centering
    \includegraphics[width=0.9\linewidth]{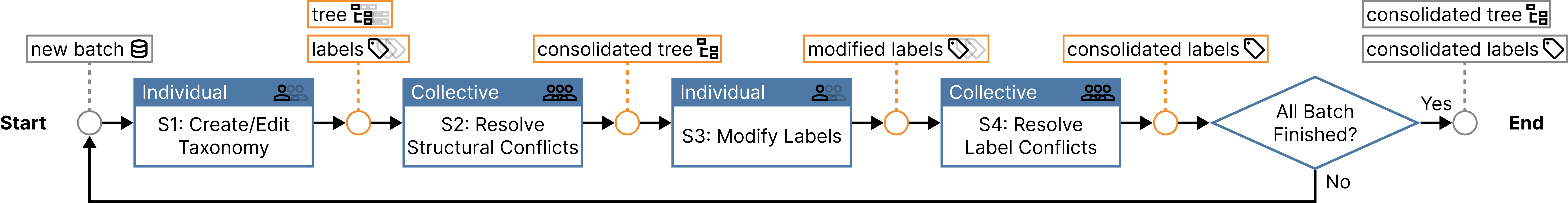}
    \caption{
        \textbf{The coding steps:}
        For each batch of images, the coders go through four steps.
        (S1) create and edit taxonomy \textit{individually},
        (S2) resolve structural conflicts \textit{collectively} with other coders,
        (S3) modify labels \textit{individually}, and
        (S4) resolve label conflicts \textit{collectively} with other coders.
        The coding is conducted iteratively for multiple image batches.
        A consolidated taxonomy tree and image labels are obtained through the coding steps.
    }
    \label{fig:coding-steps}
\end{figure*}

\section{Coding Method}
\label{sec:coding-method}

This section describes our coding protocol to operationalize the process of developing a taxonomy empirically using a collection of images.
Using the protocol, we developed an initial taxonomy of historical visualizations by carefully examining and coding \nOldVisLabeled visualization images.
Three of the authors (referred to as \textbf{\coderIndexWeiliZheng}, \textbf{\coderIndexXinyuChen}, and \textbf{\coderIndexYuZhang}) collectively conduct the coding process.
This section introduces the goal of our taxonomy (\cref{sec:goal}), the data source (\cref{sec:data-source}), the coding protocol (\cref{sec:protocol}) and its details (\cref{sec:protocol-details}), and considerations in the coding method (\cref{sec:considerations}).
Our coding process's results, analysis, and reflection are presented in \cref{sec:result-and-analysis} and \cref{sec:reflection}.
Note that the protocol and its details are generic and may be reused for other image taxonomy development scenarios, while the goal and data source are specific to our study.

\subsection{Goal}
\label{sec:goal}

Our main goal is to develop a visualization taxonomy that serves as a structured overview of the design space.
Thus, our coding process aims to group similar visualization images and assign a concise label to each group.
We do not set explicit criteria for what constitutes ``similar'' before coding and leave the judgment to the coders.
In addition, we want the labels to be concise and informative, allowing them to be used as search keywords for future integration in OldVisOnline~\cite{Zhang2024OldVisOnline}.

\subsection{Data Source}
\label{sec:data-source}

The coding is conducted on the OldVisOnline dataset~\cite{Zhang2024OldVisOnline} with \nOldVisShort images of historical visualization images published before 1950\footnote{The corpus with visualization image URLs and metadata we use is accessed at \url{https://github.com/oldvis/dataset/blob/main/dataset/output/visualizations.json} (last accessed on 2024/09/22).}.
The dataset focuses on statistical graphs.
While thematic maps may be included, maps with only cartographic representations of geographical features are excluded~\cite{Friendly2010First}.
The OldVisOnline dataset is designed to cover images that are edge cases of visualizations.
As we are interested in building a taxonomy of visualizations, we want to exclude atypical images that may not be considered visualization in common sense.
Thus, the coders may categorize some images as non-visualizations in the coding process.
The dataset was synthesized from digital libraries hosted in Europe and the United States.
Thus, its coverage is likely biased toward visualizations in these regions.

We randomly sampled four disjoint batches of images from the dataset, each consisting of 100 images.
We refer to the four batches as \textbf{B1}, \textbf{B2}, \textbf{B3}, and \textbf{B4}.

\subsection{Protocol Overview}
\label{sec:protocol}

The coding process aims to develop a taxonomy with hierarchically organized taxa and labels of the examined images.
The protocol is designed to be generic for developing taxonomy on image collections not limited to historical visualizations.
For each batch, the coders carry out the following four coding steps (\cref{fig:coding-steps}).

\begin{enumerate}[leftmargin=5.2mm]
    \item[\textbf{S1}] \textbf{Create and edit taxonomy:}
          Each coder individually creates and edits a taxonomy, given a batch of images.
          Specifically, the coder needs to iteratively edit a hierarchy of taxa and assign the images to the taxa.
          The process of assigning labels may require repeatedly reviewing the images and refining the labels.
          Each image can be assigned to multiple taxa, corresponding to multi-label classification.
          When encountering edge cases and difficulties in label assignment, coders document these circumstances and thoughts in a memo.
          \textit{Output: image labels and taxonomy tree.}

    \item[\textbf{S2}] \textbf{Resolve structural conflicts:}
          All coders collectively discuss the disagreements on the taxonomy tree, i.e., the division, naming, and definition of the taxa.
          The taxon is added to a consolidated tree when a consensus is reached.
          The coders may reserve differences if a consensus cannot be achieved.
          The discussion to resolve conflicts and the reserved differences are recorded in a memo.
          \textit{Output: consolidated taxonomy tree.}

    \item[\textbf{S3}] \textbf{Modify labels:}
          Each coder individually modifies the taxonomy tree and the image labels based on the discussion and the taxonomy tree consolidated in \textbf{S2}.
          When encountering edge cases and difficulties in label assignment, coders document these circumstances and thoughts in a memo.
          \textit{Output: modified labels.}

    \item[\textbf{S4}] \textbf{Resolve label conflicts:}
          All coders collectively go through each image whose labels differ among the coders.
          For each conflicting label, the coders justify their choices, which may involve discussing the definition of the taxon, the understanding of the image, and how the two align.
          When a conflict is resolved, the coders revise their labels accordingly.
          Similar to \textbf{S2}, the coders may reserve differences if a consensus cannot be achieved.
          The discussion to resolve conflicts and the reserved differences are recorded in a memo.
          \textit{Output: consolidated labels.}
\end{enumerate}

After a batch of images is coded, the four steps are repeated for the next batch.
When handling a new batch, the coder starts with the taxonomy tree from the previous batch.
The coded images of previous batches can be re-edited if needed.
Note that while we expect \textbf{S2} to be the primary step for resolving structural conflicts to consolidate the taxonomy tree, the coders may also edit the taxonomy tree in \textbf{S3} and \textbf{S4}.
Throughout the coding process, the coders can access external information, such as search engines and the glossary (see \cref{sec:protocol-details}).
A taxonomy is considered converged, i.e., achieves \textit{theoretical saturation}, when a new batch of images introduces no new taxa.

\subsection{Protocol Details}
\label{sec:protocol-details}

Our protocol borrows techniques from grounded theory~\cite{Corbin1990Grounded}, including \textit{constant comparisons}, \textit{memo writing}, and \textit{theoretical saturation}.
Besides, we propose compiling a \textit{glossary} to align the language use.

\textbf{Constant comparisons}
involve systematically contrasting observations to identify similarities and differences at different levels~\cite{Corbin1990Grounded}, enabling coders to iteratively refine the taxonomy at each step~\cite{Diehl2024Analysis}.

\begin{itemize}[leftmargin=*]
    \item In S1, when creating taxa, each coder \textit{compares} the key differences between images to establish boundaries between taxa.
          A coder may mistakenly assign an image to a taxon or feel uncertain about an assignment.
          When such an assignment is
          \textit{compared} with other assignments, the coder may spot inconsistencies and re-evaluate the image's features to decide whether a correction is needed.

    \item In S2, the taxonomy tree structures of different coders are collocated for \textit{comparison}.
          The discrepancies in the tree structures are algorithmically detected (see \cref{sec:taxonomy-comparison-interface}).
          The coders then \textit{compare} the conflicting taxa to resolve the structural conflicts.

    \item In S3, when modifying labels, each coder \textit{compares} different taxa options (including newly proposed or refined taxa from S2) for an image to make adjustments to its labels.

    \item In S4, different labels assigned to the same image by the different coders are collocated for \textit{comparison}.
          The label discrepancies are algorithmically detected (see \cref{sec:taxonomy-comparison-interface}).
          The coders then \textit{compare} the conflicting labels to resolve the label conflicts.
\end{itemize}

\textbf{Memo writing}
helps coders track edge cases, questions, reserved differences, and changes~\cite{Corbin1990Grounded}.
Coders may pin down the rationale behind label decisions, reflect on questions and reserved differences, and track changes to the taxonomy tree and changes of each taxa's names and definitions, ultimately leading to a consolidated taxonomy.
In our protocol, coders employ memo writing in all steps.

\begin{itemize}[leftmargin=*]
    \item In the individual editing stages (S1 and S3), when naming a new taxon, coders write \textit{memos} on its definition and decision criteria.
          For special cases (e.g., an image hard to categorize), coders write \textit{memos}.

    \item In the conflict resolution stages (S2 and S4), coders write \textit{memos} about the causes for their disagreements and how they reached an agreement through discussion.
\end{itemize}

\textbf{Theoretical saturation}
occurs when further data collection no longer yields new insights, signaling that the researcher has achieved empirical confidence in the taxonomy~\cite{Glaser1967Discovery}.
At the end of each batch, we check whether the taxonomy tree has stabilized to determine whether the coding iteration should stop.
The saturation checking validates that the taxonomy generalizes to newly seen images.

\textbf{Glossary} is used to improve the consistency of language use by different coders.
A visual design may have multiple names.
Prior knowledge of the coders may affect how they refer to and name visual designs.
Thus, we use a glossary of terms referring to visual designs to ground the coders in a shared context and language.
The terms are extracted from \nGlossaryPaper papers related to visualization classification schemes that use terms referring to visual designs as taxa.
\iflabelexists{sec:glossary-of-terms}{\Cref{sec:glossary-of-terms}}{The supplemental material titled ``Glossary of Terms''}
describes the terms we obtained from the literature.
The glossary is a reference that the coders may access before and during coding to inform the naming of taxa.
Before coding started, all the coders went through the glossary to familiarize themselves with the terms and discussed with each other to align on the interpretation of the glossary.
The coders may also search for or come up with names outside the collection if they come across cases that do not match any of the terms in the glossary.
Note that consulting existing terms relates to our goal of using the taxa names as future search keywords.
A related practice in the literature was by Fujishiro et al.~\cite{Fujishiro2000GADGET_IV}, where they used a thesaurus to normalize the naming of visualization tasks in collected papers.

\subsection{Protocol Considerations}
\label{sec:considerations}

This section discusses the considerations in the protocol design.

\textbf{On the connection with grounded theory:}
Grounded theory, commonly used in social sciences, has a systematic workflow for theory generation grounded on emergent data~\cite{Corbin1990Grounded}.
Typically, its final generated theory aims to understand and explain phenomena through a systematic explanation or framework from concepts and their relationships.
In our scenario, the full-fledged practice of grounded theory is unnecessary.
Our objective in this work is more inclined toward utilizing tree-structured relationships between taxonomies to aid in understanding and presenting the design of historical visualization rather than establishing a theory to explain historical phenomena.
As our scenario concerns descriptive instead of explanatory framework, we do not directly use grounded theory but borrow techniques from it.
Our protocol can be seen as a partial adoption of the coding steps in grounded theory, which is more lightweight.

\textbf{On multiple coders:}
We involve multiple coders for two reasons:

\begin{itemize}[leftmargin=*]
    \item \textbf{Fix mistakes:} Examining the conflicts among coders helps fix coding mistakes caused by carelessness, including inconsistent label assignments and language use.

    \item \textbf{Reveal perspectives:} More importantly, conflicts may reveal different perspectives on the visualization design space for debate.
\end{itemize}

While our protocol encourages coders to resolve conflicts, it allows coders to reserve differences.
As coding is a creative task without ground truth, the coders may hold different opinions, particularly on edge cases of each taxon.
For disagreements that cannot be resolved, the coders may set aside the disputes and record their arguments.
They can return to the discussion when finding relevant new cases and evidence.
Additionally, the consistent disagreements among coders are interesting cases that we want to examine.

Note that our purpose for involving multiple coders differs from the setup of data labeling.
In data labeling, particularly crowdsourcing data labels, it is common to involve multiple labelers and adopt the most popular labels through voting.
The voting process implicitly assumes a ``ground truth'' that the majority can identify.
By comparison, when categorizing historical visualizations, we do not expect a ``ground truth'' taxonomy to exist, nor do we expect the majority to be correct.

\textbf{On splitting conflict resolution into two steps:}
We split the conflict resolution into two steps: resolving structural conflicts and resolving label conflicts.
An alternative setup is to merge S2, S3, and S4 into one step, in which case the coders resolve all conflicts in one go.
Our rationale for splitting the steps is that when the taxonomy trees of different coders are different, the labels assigned to the images systematically differ.
Thus, we regard it more efficient to resolve the structural conflicts before resolving the label conflicts.

\textbf{On the batch division:}
We split the dataset into small batches, each with 100 images, and iterated the coding process for each batch.
The batch division enables coders to frequently align and consolidate their taxonomies through discussion.

\textbf{On the inductive approach:}
In our practice, the coders started from scratch to develop a new taxonomy inductively.
An alternative deductive approach is to start with an existing taxonomy, assign the images to the taxa, and refine the taxonomy if needed.
While our protocol does not prohibit the deductive approach, we did not start with an existing taxonomy to eliminate the potential anchoring effect~\cite{Tversky1974Judgment}.

    \begin{figure*}[!htbp]
    \centering
    \includegraphics[width=0.87\linewidth]{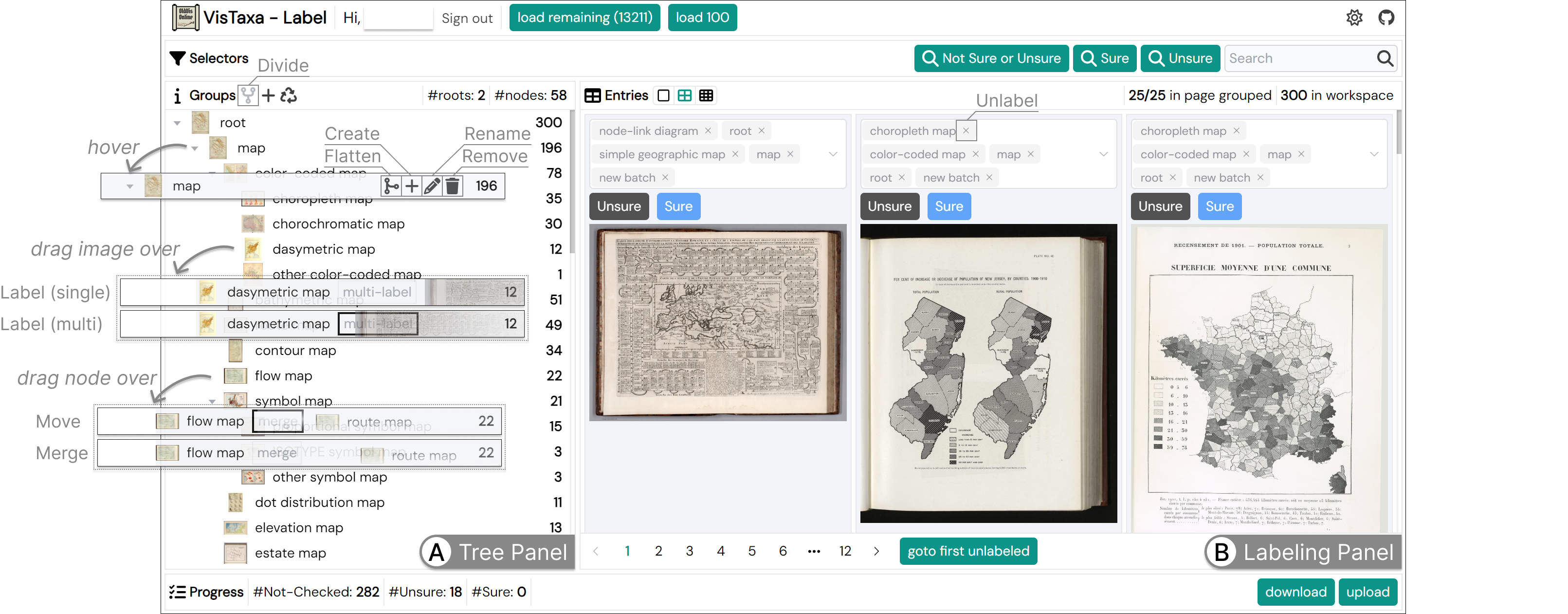}
    \caption{
        \textbf{The \systemName interface for taxonomy labeling:}
        (A) The tree panel shows the taxonomy tree.
        Within the tree panel, the coder can edit the taxonomy tree by operators such as creating/renaming/dividing/flattening/removing/moving a taxon node and merging two taxon nodes.
        The coder may drag the image and drop it onto a taxon node to assign the image to the taxon.
        (B) The labeling panel shows the taxonomy labels.
        Within the labeling panel, the coder can edit the taxonomy label of each image.
    }
    \label{fig:interface-taxonomy-labeling}
\end{figure*}

\section{\systemName: A System for Image Taxonomy Development}
\label{sec:system}

This section introduces the \systemName system we developed for image taxonomy development.
To serve the coding steps introduced in \cref{sec:protocol}, \systemName consists of two interfaces:

\begin{itemize}[leftmargin=*]
    \item The \textbf{taxonomy labeling interface} (see \cref{fig:interface-taxonomy-labeling}) serves S1 (Create/Edit Taxonomy), S3 (Modify Labels), and S4 (Resolve Label Conflicts) for coders to \textit{edit} the taxonomy tree and the image labels (\cref{sec:taxonomy-labeling-interface}).
    \item The \textbf{taxonomy comparison interface} (see \cref{fig:interface-taxonomy-comparison}) serves S2 (Resolve Structural Conflicts) and S4 (Resolve Label Conflicts) for coders to \textit{compare} taxonomy trees and image labels (\cref{sec:taxonomy-comparison-interface}).
\end{itemize}

\subsection{Taxonomy Labeling Interface}
\label{sec:taxonomy-labeling-interface}

The following first formalizes the image taxonomy labeling task to clarify the notation and inform the functions to be supported in an image taxonomy labeling interface.
We then introduce the operators supported in our taxonomy labeling interface and the machine assistance employed for efficient taxonomy labeling.

\subsubsection{Notation and Task Formalization}
\label{sec:notation-and-task-formalization}

Image taxonomy labeling can be divided into two subtasks as follows.

\begin{itemize}[leftmargin=*]
    \item \textbf{Tree editing task:} The coder needs to build a tree that stores the taxonomy structure.
          Specifically, the coder needs to define a rooted tree $T$ with nodes $V = \{ v_i \}_{i=1}^n$ and edges $E = \{ e_i \}_{i=1}^{n-1}$.
          Each node $v_i$ is associated with a name $name(v_i)$.

    \item \textbf{Label editing task:} The coder needs to assign each image to at least one leaf node in the tree.
          Specifically, given a set of visualization images $D = \{ d_i \}_{i=1}^m$, the coder needs to assign each image $d_i$ to a set of leaf nodes $V_i = \{ v_{o_j} \}_{j=1}^{n_i} \subseteq V$.
\end{itemize}

We allow an image to be assigned to multiple taxa, corresponding to multi-label classification.
The coder may use multiple labels to handle composite visualizations.

\subsubsection{Operators}

The following lists the operators supported in the taxonomy labeling interface (\cref{fig:interface-taxonomy-labeling}).
We describe each operator's impact on the taxonomy tree and image labels.

\begin{itemize}[leftmargin=*]
    \item \textbf{Create} a taxon:
          Appends a node $v$ to a parent node $v'$ in the tree.
          If $v$ is the only child of $v'$, an additional $v_{ungrouped}$ is appended to $v'$.
          If $v$ is the only child of $v'$, the images in $v'$ are assigned to $v_{ungrouped}$ to ensure the assigned labels correspond to a leaf node.

    \item \textbf{Divide} a taxon:
          Divides a node $v$ into multiple nodes $v_1$, $v_2$, ..., $v_n$ (using the clustering process in \cref{sec:machine-assistance}).
          The images in $v$ are assigned to the new nodes $v_1$, $v_2$, ..., $v_n$.

    \item \textbf{Flatten} a taxon:
          Removes the children $v_1$, $v_2$, ..., $v_n$ of node $v$.
          The images in $v_1$, $v_2$, ..., $v_n$ are assigned to $v$.
          A flatten operation can be used to undo a divide operation.

    \item \textbf{Merge} two taxa:
          Merges two nodes $v_1$ and $v_2$ into a single node $v_1$.
          The images in $v_2$ are assigned to $v_1$.

    \item \textbf{Move} a taxon:
          Moves a node $v$ from parent node $v_1$ to a new parent node $v_2$.

    \item \textbf{Rename} a taxon:
          Change the name $name(v)$ of a node $v$.

    \item \textbf{Remove} a taxon:
          Remove a node $v$ and its descendants.
          The images assigned to $v$ and its descendants are unassigned.

    \item \textbf{Label} an image:
          Assign an image $d$ to a node $c$.

    \item \textbf{Unlabel} an image:
          Unassign an image $d$ from a node $c$.
\end{itemize}

In addition to the above operators for taxonomy editing, \systemName provides other functions for handling images incrementally and reviewing the annotations.
In the scenarios where the image set is large or where the image set is incrementally enlarged, \systemName supports incrementally loading images in batches.
The newly loaded images are assigned to an \taxon{ungrouped} taxon by default.
The coder may review the annotations by image filtering and searching.
The coder can filter images by a taxon or search images by keywords or UUIDs.

\textbf{Usage in the coding process:}
The taxonomy comparison interface serves S1, S3, and S4.
In the three steps, the coders may use the operators to edit the taxonomy tree and the image labels.

\begin{figure}[!htbp]
    \centering
    \includegraphics[width=1\linewidth]{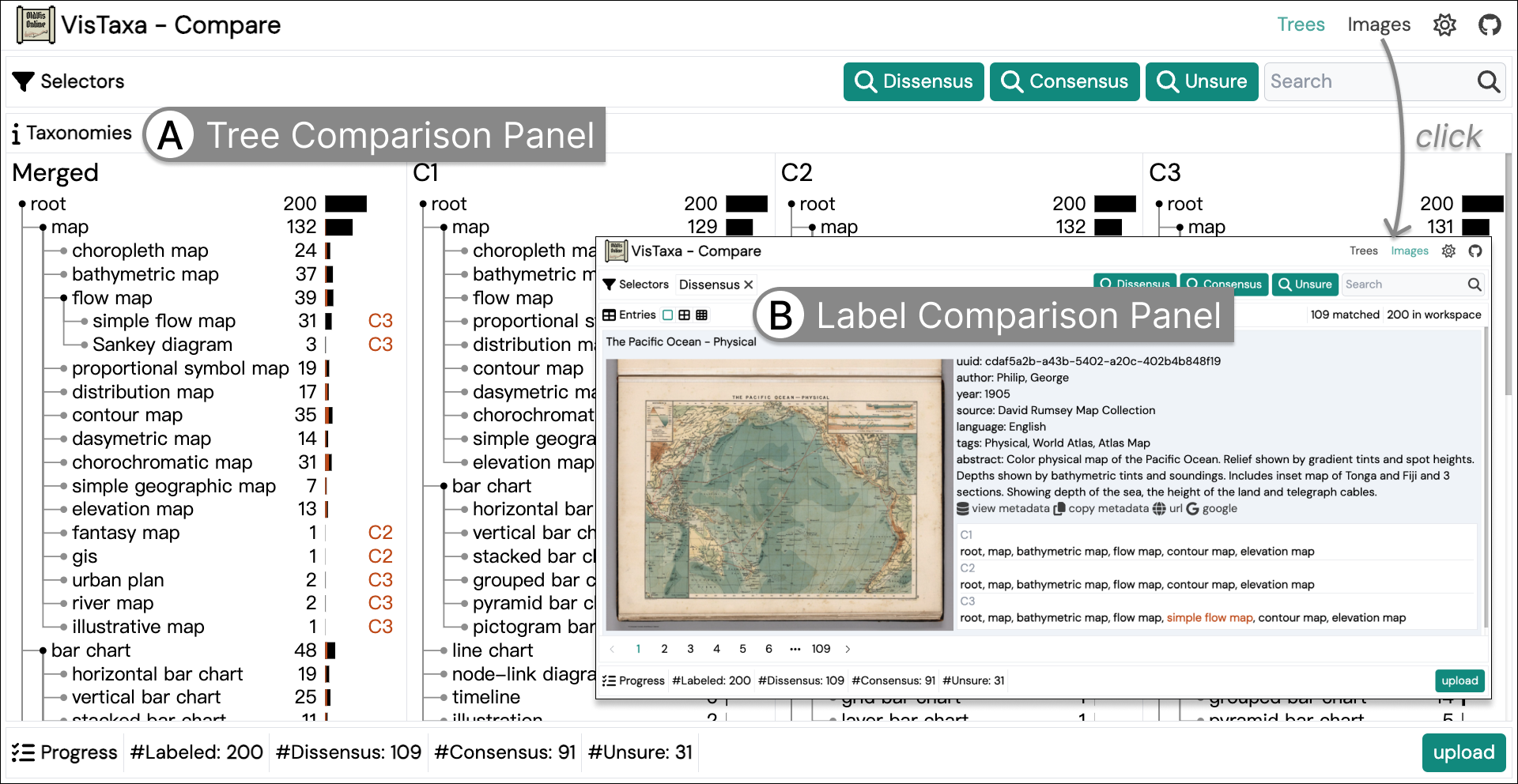}
    \caption{
        \textbf{The \systemName interface for taxonomy comparison:}
        (A) The tree comparison panel shows the taxonomy tree constructed by each coder (\coderIndexWeiliZheng, \coderIndexXinyuChen, and \coderIndexYuZhang) and the merged taxonomy tree.
        (B) The label comparison panel shows the images and their assigned taxa.
    }
    \label{fig:interface-taxonomy-comparison}
\end{figure}

\subsubsection{Machine Assistance in Taxonomy Labeling}
\label{sec:machine-assistance}

When creating and refining a taxonomy, a common need is to split a taxon into multiple taxa.
It corresponds to dividing a concept at a coarse granularity (e.g., \taxon{bar chart}) into multiple concepts at a finer granularity (e.g., \taxon{stacked bar chart} and \taxon{grouped bar chart}).
Manually splitting a taxon may involve creating new taxa and assigning images to the new taxa one by one, which is time-consuming.
To support efficient splitting of taxa, \systemName employs two types of machine assistance: \textit{clustering} and \textit{semantic cluster overview}.

\textbf{Clustering:}
A clustering process groups images.
For each image, we use CLIP ViT-B/32~\cite{Radford2021Learning} to extract a 512-dimension feature vector.
When the user applies the divide operator to a taxon in \systemName, the system automatically clusters the images in the taxon.
Given a taxon with $n$ images, we use KMeans to group these images into $\left \lfloor{\sqrt{n}}\right \rfloor$ clusters.
Each cluster corresponds to an algorithmically created taxon.
The coder can then refine the taxon with the editing operators.

\textbf{Semantic cluster overview:}
For each cluster shown as a node in the taxonomy tree, \systemName shows a representative image and a textual overview.
The image with the embedding closest to the centroid is chosen as the representative image.
For each image, we generate a textual description by prompting BLIP-2 OPT-2.7B~\cite{Li2023BLIP} with \inlinecodesmall{``Question: what is the type of the visualization? Answer:''}.
We postprocess model outputs with rules to trim the part useless for understanding the image content, such as ``it is a''.
Empty string outputs are replaced with ``unknown''.
For each cluster node, we show the textual description of its representative image.

\subsection{Taxonomy Comparison Interface}
\label{sec:taxonomy-comparison-interface}

The user can upload multiple pieces of taxonomy labeling results exported from the taxonomy labeling interface to the taxonomy comparison interface.
In this way, the user can compare the taxonomy trees and the image labels among different coders.

\textbf{Taxonomy tree comparison:}
The tree comparison panel (\cref{fig:interface-taxonomy-comparison}(A)) shows the taxonomy tree constructed by each coder and the merged taxonomy tree.
The merged tree is a union of each coder's tree.
In the merging, the identity of a node is determined not by its own name but by the list of node names on the path from \taxon{root} to the node, including the node itself.
(The following refers to such case as ``the node identity is defined by the path from the root''.)
Two types of discrepancies among coders are highlighted in the merged tree.
\begin{itemize}[leftmargin=*]
    \item When a taxon is not created by all coders, it is followed by a list of coders who have created this taxon.
    \item When a taxon is created by all coders but assigned different images, the portion of images not assigned by all coders is highlighted.
\end{itemize}

\textbf{Image label comparison:}
The label comparison panel (\cref{fig:interface-taxonomy-comparison}(B)) shows the images and the taxa they are assigned by each code, with discrepancies highlighted.
The \textit{Dissensus} selector selects the images with discrepancies.
The \textit{Unsure} selector selects the images marked unsure by at least one coder.

\textbf{Usage in the coding process:}
The taxonomy comparison interface serves S2 and S4.
In S2, the coders are expected to examine the discrepancies in the taxonomy trees mainly in the tree comparison panel (\cref{fig:interface-taxonomy-comparison}(A)) to motivate discussions.
In S4, the coders are expected to focus on the label comparison panel (\cref{fig:interface-taxonomy-comparison}(B)).

    \begin{figure}[!htbp]
    \centering
    \includegraphics[width=1.0\linewidth]{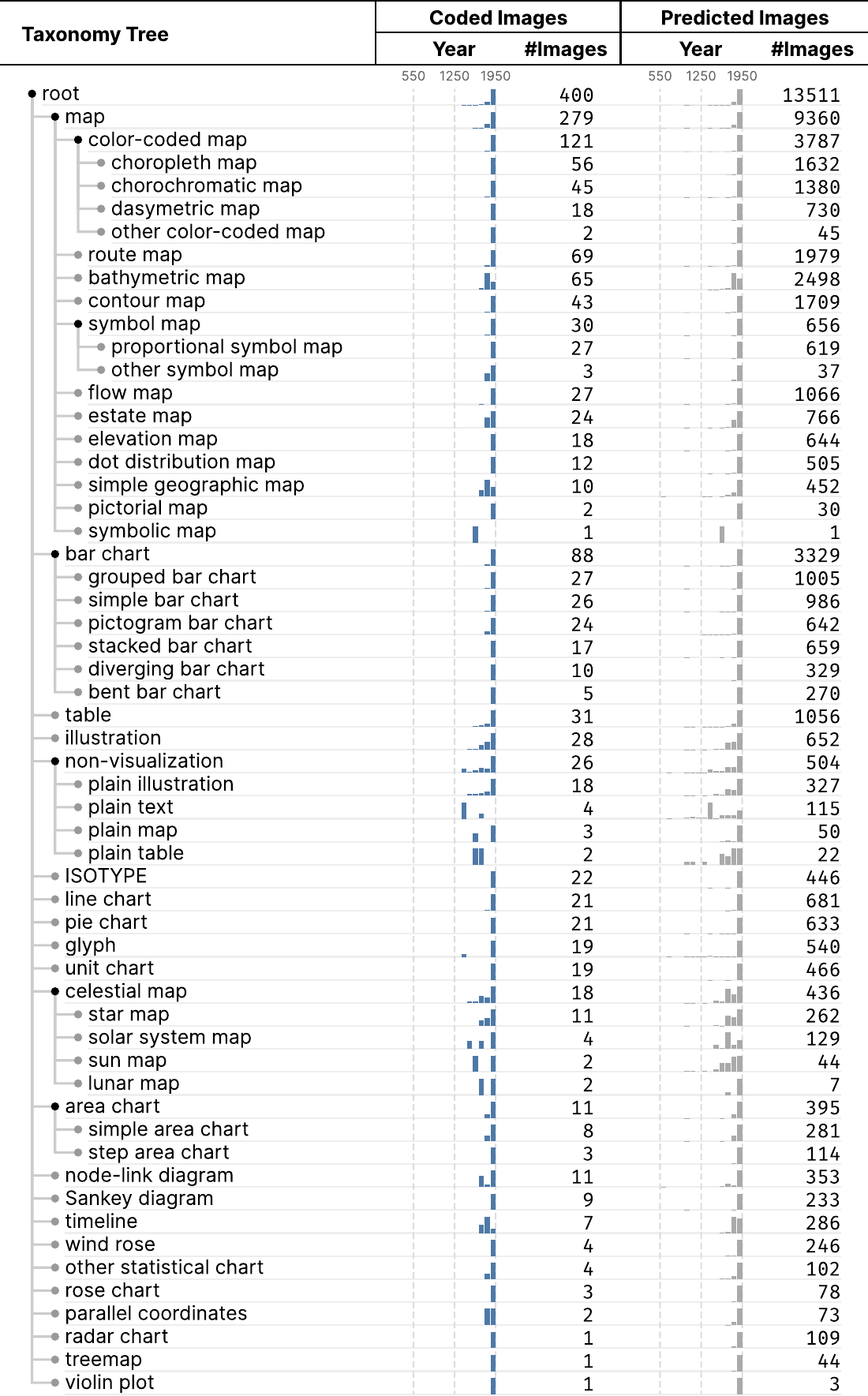}
    \caption{
        \textbf{The final taxonomy:}
        The resulting taxonomy has \nTaxaFinalAllVis taxa (excluding \taxon{root}, \taxon{non-visualization}, and the subcategories of \taxon{non-visualization}).
        \nTaxaFinalFirstLevelVis taxa are at the first level.
        (\taxon{root} is considered as at the zeroth level.)
        \nTaxaFinalLeafVis taxa are leaf nodes.
        The ``\#Images'' column in ``Coded Images'' shows the number of images assigned to each taxon by the coders.
        The ``Year'' column shows the distribution of the publish year of image in the taxon.
        (The publish year is from digital libraries used in the OldVisOnline dataset and can be inaccurate.)
        Note that an image can be assigned to multiple taxa.
        The OldVisOnline dataset contains \nOldVis images.
        For the \nOldVisPredicted not yet human-coded, we use similarity-based matching (\cref{sec:predicting-taxonomy-labels}) to predict taxonomy labels.
        The ``Predicted Images'' column shows the distribution of predicted images together with the \nOldVisLabeled labeled image.
        (Among the \nOldVisPredicted images, \nOldVisPublishYearUnknown images with unknown publish year are excluded from the histogram of publish years.)
        The prediction is imperfect and should be interpreted with caution.
    }
    \label{fig:taxonomy}
\end{figure}

\section{Result and Analysis}
\label{sec:result-and-analysis}

This section presents the taxonomy we developed by coding \nOldVisLabeled historical visualizations.
With the labels of \nOldVisLabeled images, we predict the labels of \nOldVisShort historical visualizations to obtain a coarse overview of the distribution of visual designs in historical visualizations.
We analyze the distribution patterns and compare our taxonomy with existing schemes.

\subsection{The Final Taxonomy}
\label{sec:the-final-taxonomy}

\Cref{fig:taxonomy} shows the final taxonomy with \nTaxaFinalAllVis taxa (excluding \taxon{root}, \taxon{non-visualization}, and the subcategories of \taxon{non-visualization}).
The final taxonomy tree and image labels are generated by merging the three coders' taxonomy trees and labels assigned to the \nOldVisLabeled images at the end of the coding process.

\begin{itemize}[leftmargin=*]
    \item \textbf{Merged taxonomy tree:}
          The taxonomy trees are merged by majority voting\footnote{
              The merging strategy here differs from that in \cref{sec:taxonomy-comparison-interface}.
              Here, we use majority voting to resolve conflicts, while \cref{sec:taxonomy-comparison-interface} uses union to expose conflicts.
          }.
          A node is kept in the merged tree if it exists in more than half of the coders' taxonomy tree.
          In the voting, the node identity is defined by the path from the root.

    \item \textbf{Merged image labels:}
          The image labels are merged by majority voting.
          We assign a taxonomy label to an image if more than half of the coders have assigned the label to the image.
          In the voting, the label identity is defined by the path from the root.
\end{itemize}

\iflabelexists{fig:examples}{\Cref{fig:examples}}{The figure in supplemental material titled ``Example images in each leaf taxon''} shows example images of leaf taxa.
\iflabelexists{tab:taxon-definitions}{\Cref{tab:taxon-definitions}}{The table in supplemental material titled ``Taxon definitions''} describes the definition of each taxon.

\begin{table}[!htbp]
    \caption{
        The performance of taxonomy label prediction with similarity-based matching and zero-shot classification.
        Exact match ratio (denoted $Match$) and Jaccard score (denoted $Jaccard$) are the metrics.
        $D = 1$ denotes evaluating with the labels at the first level of the taxonomy tree.
    }
    \label{tab:predicting-taxonomy-labels}
    \centering
    \scriptsize
    \scalebox{0.83}{
        \begin{tabular}{lrrrr}
            \toprule
            \textbf{Method}           & $Match$ & $Jaccard$ & $Match~(D = 1)$ & $Jaccard~(D = 1)$ \\
            \midrule
            similarity-based matching & 39.75\% & 0.619     & 65.50\%         & 0.767             \\
            zero-shot classification  & 9.00\%  & 0.291     & 53.25\%         & 0.617             \\
            \bottomrule
        \end{tabular}
    }
\end{table}

\subsection{Predicting Taxonomy Labels}
\label{sec:predicting-taxonomy-labels}

Through the coding process, we have obtained the taxonomy labels of \nOldVisLabeled images among the \nOldVisShort images of the OldVisOnline dataset~\cite{Zhang2024OldVisOnline}.
This section describes predicting the taxonomy labels for the remaining \nOldVisPredicted unlabeled images.
The prediction is a multi-label classification task.
We have trialed two prediction methods:

\textbf{Similarity-based matching}
predicts the taxonomy label of an image as the taxonomy label of the most similar labeled image.
We match the image based on the cosine similarity between the image embeddings extracted by CLIP ViT-B/32~\cite{Radford2021Learning}.

\textbf{Zero-shot classification}
predicts the taxonomy label of an image given the candidate labels.
We use CLIP ViT-B/32 to classify with the names of leaf taxa being candidate labels.
We assign an image to the label with the highest probability and all other labels with probabilities no less than $0.3$.
When an image is predicted to belong to a leaf taxon, it is also assigned to the leaf taxon's ancestor taxa.

We evaluate the performance of each method on the \nOldVisLabeled labeled images with two metrics:

\begin{itemize}[leftmargin=*]
    \item \textbf{Exact match ratio:}
          The rate of images where the predicted labels exactly match the merged coder labels.

    \item \textbf{Jaccard score:}
          For an image with predicted label set $A$ (excluding \taxon{root}) and the label set $B$ generated by coders (excluding \taxon{root}), the Jaccard similarity is defined as $\frac{|A \cap B|}{|A \cup B|}$.
          We average the Jaccard similarity over all images.
\end{itemize}

We use leave-one-out cross-validation when evaluating similarity-based matching.
For each metric, we consider two setups: evaluate with the full taxonomy labels and evaluate with the taxonomy label at the first level of the tree (denoted as $D = 1$).

\Cref{tab:predicting-taxonomy-labels} shows that similarity-based matching significantly outperforms zero-shot classification.
The Jaccard Score of similarity-based matching is $0.619$ while that of zero-shot classification is $0.290$.
We interpret that \textit{taxon names are less informative than labeled images} for the model to categorize images.

Thus, we choose similarity-based matching to predict the taxonomy labels of the \nOldVisPredicted unlabeled images.
The ``Predicted Images'' column in \cref{fig:taxonomy} shows the number of predicted images of each taxon and their temporal distribution.

\subsection{Distribution Analysis}
\label{sec:distribution-analysis}

The taxonomy categorizes images, allowing analysis of the distribution of specific image types.
The following analyzes distribution patterns that \cref{fig:taxonomy} reveals\footnote{
    Our analysis is for inspiring further investigation.
    It should not be seen as a definitive conclusion because of potential imprecisions introduced by dataset limitations, potential coder errors, and model prediction errors.
}.

\textbf{Comparison with contemporary frequent chart types:}
Battle et al.~\cite{Battle2018Beagle} found \taxon{geographic map}, \taxon{line chart}, \taxon{bar chart}, and \taxon{scatter chart} to be the most common chart types in SVG-based visualizations on the web.
\Cref{fig:taxonomy} shows that \textit{the most frequent types of historical visualizations are mostly consistent with contemporary visualizations on the web}: \taxon{map}, \taxon{bar chart}, and \taxon{line chart} are among the most frequent taxa.
However, \taxon{scatter chart} is absent in our taxonomy.
A possible explanation is that scatterplot was invented later than the other contemporary frequent chart types and thus has fewer occurrences in historical visualizations.
Friendly and Denis~\cite{Friendly2005Early} attribute the earliest known scatterplot to Herschel in 1833~\cite{Herschel1833Investigation}.
By comparison, the invention of bar chart and line chart is attributed~\cite{Friendly2001Milestones} to Playfair in 1786~\cite{Playfair1786Commercial}, and maps have been used for thousands of years~\cite{Friendly2001Milestones}.

\textbf{Temporal distribution and outlier:}
As the temporal distribution of images in \taxon{root} shows, the number of images increases over time.
Interestingly, we observe several counterexamples that do not follow this trend.
A notable example is the temporal distribution of \taxon{bathymetric map}, which peaks around 1800 CE and then decreases over time.
An interpretation may be that the colonial expansion of European countries in the 19th century increased the use of bathymetric data.

\textbf{``Futuristic'' invention:}
We observe that some taxa that may be considered as ``modern visualization'' actually appeared early.
An \href{https://www.davidrumsey.com/luna/servlet/detail/RUMSEY~8~1~1499~160037}{example} is that the prototype of parallel-coordinate design appeared at least as early as 1849 to represent hydrological information~\cite{Berghaus1849Vermischtes}.

\subsection{Comparison with Existing Schemes}
\label{sec:comparison-with-existing-schemes}

\Cref{tab:taxonomies} compares high-level characteristics of our taxonomy with existing schemes.
This section compares the content of \scheme{Ours} with three schemes: \scheme{Lohse1994}~\cite{Lohse1994Classification}, \scheme{Friendly2001}~\cite{Friendly2001Milestones}, and \scheme{Chen2024}~\cite{Chen2024Image}.

\subsubsection{Comparison with a Visual Representation Classification}

\scheme{Lohse1994} is a classification of visual representations with 8 categories: \taxon{graph}, \taxon{table}, \taxon{time charts}, \taxon{network chart}, \taxon{map}, \taxon{cartogram}, \taxon{icons}, and \taxon{photo-realistic picture}~\cite{Lohse1994Classification}.
\taxon{table} has 2 subcategories: \taxon{numeric table} and \taxon{graphic table}.
\taxon{network chart} has 2 subcategories: \taxon{structure diagram} and \taxon{process diagram}.

\textbf{Naming convention:}
\scheme{Lohse1994} and \scheme{Ours} both use common visual representation names as category names (e.g., \taxon{table} and \taxon{map}).

\textbf{Scope:}
Overall, \scheme{Lohse1994} has a broader scope than \scheme{Ours}, which covers visual representations that are largely irrelevant to visualization (e.g., \taxon{icons} and \taxon{photo-realistic picture}).
This difference reflects the different scopes of the coded corpus.

\textbf{Organization structure:}
In \scheme{Lohse1994}, \taxon{map} and \taxon{cartogram} are at the same level.
By comparison, in \scheme{Ours}, map types are categorized under \taxon{map}.
It seems more reasonable to let \taxon{cartogram} be a subcategory of \taxon{map}.
This artifact in \scheme{Lohse1994} may be attributed to its method that relies on algorithmic voting instead of discussion to resolve disagreements.
Specifically, the multiple participants' grouping of images is used to construct a similarity matrix.
The similarity of two categories is $N$ when the two categories are grouped in $N$ subjects' lowest level grouping.
Complete linkage hierarchical clustering is then applied to the similarity matrix to generate the hierarchical classification.

\subsubsection{Comparison with a Visualization Milestone Categorization}

\scheme{Friendly2001}\footnote{The categorization is on page 41 of the \href{http://euclid.psych.yorku.ca/SCS/Gallery/milestone/milestone.pdf}{PDF version} of the Milestones project~\cite{Friendly2001Milestones} (last accessed on 2024/11/15).}
is a categorization of milestones (i.e., major inventions and progress) in the history of thematic cartography, statistical graphics, and data visualization.
The categorization has two facets: content (e.g., \taxon{astronomy} and \taxon{commerce}) and form (e.g., \taxon{chart} and \taxon{curve}).
As there are hundreds of categories in \scheme{Friendly2001}, we do not list each for space limit.

\textbf{Scope:}
\scheme{Friendly2001} has a broader scope than \scheme{Ours}.
Milestones in the visualization history concerned in \scheme{Friendly2001} are not limited to the visualization field.
Thus, taxa in \scheme{Friendly2001} includes inventions in relevant fields, such as \taxon{least squares} in statistics.
Additionally, \scheme{Friendly2001} covers milestones in late modern history, such as \taxon{chernoff face}, and interaction techniques, such as \taxon{table lens}.
These instances are outside the scope of \scheme{Ours}.

\textbf{Organization structure:}
\scheme{Friendly2001} is a faceted scheme.
Its ``content'' facet is outside the scope of \scheme{Ours}, while its ``form'' facet is relevant to \scheme{Ours}.
In the following, we primarily focus on \scheme{Friendly2001}'s ``form'' facet for relevance.
The ``form'' facet has 16 categories (each with various subcategories): \taxon{apparatus}, \taxon{chart}, \taxon{coordinates}, \taxon{curve}, \taxon{diagram}, \taxon{graph}, \taxon{grid}, \taxon{history}, \taxon{line}, \taxon{map}, \taxon{mathematics}, \taxon{pattern}, \taxon{perspective}, \taxon{plot}, \taxon{table}, and \taxon{visual}.

\textbf{Goal:}
\scheme{Friendly2001} mainly serves as an index for readers to look up historical references.
For this goal, \scheme{Friendly2001} intentionally includes synonyms to make searching easier for readers.
For example, \taxon{scatter graph}, \taxon{scatter plot}, \taxon{scattergraph}, and \taxon{scatterplot} are all included as subcategories of \taxon{graph}.
By comparison, \scheme{Ours} serves as a concise overview of the design space and thus contains no synonyms.

\subsubsection{Comparison with an IEEE VIS Image Typology}

\scheme{Chen2024} is a typology of images in IEEE VIS papers with 10 categories: \taxon{surface/volume}, \taxon{line}, \taxon{point}, \taxon{bar}, \taxon{color/greyscale}, \taxon{node-link}, \taxon{area}, \taxon{grid}, \taxon{glyph}, and \taxon{text}~\cite{Chen2022Not,Chen2024Image}.

\textbf{Naming convention:}
\scheme{Chen2024} is inspired by Bertin's semiology of graphics~\cite{Bertin1981Graphics}, and thus the category names relate to visual mark names (e.g., \taxon{bar}).
In contrast, the category names in \scheme{Ours} relate to chart names (e.g., \taxon{bar chart}), as \scheme{Ours} are developed to serve as search keywords and are thus more comprehensible to non-specialists.
Note that Chen et al. discussed that using technique names as categories may not be scalable for categorizing VIS images, as VIS papers are intended to introduce new techniques frequently.
In the case of historical visualizations, while they feature diverse designs, the number of unique techniques, in terms of chart types, is relatively limited.
It is thus manageable to categorize historical visualizations by chart names.

\textbf{Scope:}
\scheme{Chen2024} has a category \taxon{surface/volume} which includes various 3D scientific visualizations.
Such designs hardly appear in historical visualizations before the computer era, and thus \scheme{Ours} does not have such a category.
Additionally, \taxon{map} is a well-developed category in \scheme{Ours}, with various subcategories.
By comparison, maps are categorized under the \taxon{area} category in \scheme{Chen2024} with other area representations such as area charts and pie charts.
The relatively low presence of maps in VIS publications may reflect that cartography and visualization are largely two separate fields nowadays.

    \section{Reflection on the Coding Process}
\label{sec:reflection}

This section examines the statistics on the coding process and introduces observations on the coding criteria of coders.

\subsection{Statistics of the Coding Process}
\label{sec:statistics}

This section analyzes the coding process with three metrics, the number of nodes, time cost, and level of agreement, as \cref{tab:coding-statistics} shows.

\textbf{Number of nodes:}
The number of nodes in the merged taxonomy tree ($n_m$) increases through the steps from 7 at B1S1 (abbreviation for Batch 1 Step 1) to 56 at B4S4.
19 additional nodes were added throughout the four steps of Batch 2, only 7 additional nodes were added for Batch 3, while the number of nodes stabilized for Batch 4.
In summary, \textit{the number of nodes converges}.
The number of changed nodes ($\Delta n_m$) decreases through the steps, also suggesting saturation.

\begin{table}[!htbp]
    \caption{
        \textbf{The number of nodes, time cost, and agreement level among coders at each coding step:}
        $BiSj$ is the $j$-th step (of the four steps in \cref{sec:protocol}) of the $i$-th batch (of the four image batches).
        $ n_1, n_2, n_3 $ are the number of nodes (excluding \taxon{root}) in the taxonomy tree of each coder.
        $ n_m $ is the number of nodes in the merged taxonomy tree by majority voting (see \cref{sec:the-final-taxonomy} for the merging process).
        $ \Delta n_m $ is the number of nodes in the symmetric difference between nodes in the merged taxonomy tree of the current and the previous step.
        (The node identity is defined by the path from the root.)
        $ t_1, t_2, t_3 $ are the time cost (in hours) for coder \coderIndexWeiliZheng, \coderIndexXinyuChen, and \coderIndexYuZhang.
        $\overline{t}$ is the average time cost value of the three coders.
        ${Match}$ is the exact match ratio (i.e., the rate of images with the same label names from all coders).
        $\overline{Jaccard}$ is the Jaccard score pairwise averaged among coders.
        $Node~IoU$ is the intersection over union of the sets of nodes from all coders.
        (The node identity is determined the same as that for $ \Delta n_m $.)
        Step 1 and step 3 are conducted by each coder individually, while step 2 and step 4 are conducted by all coders collectively.
        For step 2 and step 4, we report the time cost of the collective discussion.
        Step 2 is a discussion step that does not involve taxonomy editing, and thus, we use a dash to denote the absence of values.
    }
    \setlength{\aboverulesep}{0pt}
    \setlength{\belowrulesep}{0pt}
    \setlength{\extrarowheight}{1.5pt}
    \label{tab:coding-statistics}
    \centering
    \scriptsize
    \scalebox{0.78}{
        \setlength{\tabcolsep}{0.4em}
        \begin{tabular}{l|rrrrr|rrrr|rrr}
            \toprule
            \multirow{2}[0]{*}{\textbf{Step}} & \multicolumn{5}{c|}{\textbf{\#Nodes}} & \multicolumn{4}{c|}{\textbf{Time Cost (h)}} & \multicolumn{3}{c}{\textbf{Agreement}}                                                                                                                                                                                                                                                                                                                                                           \\\cline{2-13}
                                              & \multicolumn{1}{r}{\textbf{$n_1$}}    & \multicolumn{1}{r}{\textbf{$n_2$}}          & \multicolumn{1}{r}{\textbf{$n_3$}}     & \multicolumn{1}{r}{\textbf{$n_m$}} & \multicolumn{1}{r|}{\textbf{$\Delta n_m$}} & \multicolumn{1}{r}{\textbf{$t_1$}} & \multicolumn{1}{r}{\textbf{$t_2$}} & \multicolumn{1}{r}{\textbf{$t_3$}} & \multicolumn{1}{r|}{\textbf{$\overline{t}$}} & \multicolumn{1}{r}{$Match$} & \multicolumn{1}{r}{$\overline{Jaccard}$} & \multicolumn{1}{r}{$Node~IoU$} \\
            \midrule
            B1S1                              & $11$                                  & $27$                                        & $19$                                   & $7$                                & $7$                                        & $0.48$                             & $0.97$                             & $0.76$                             & $0.73$                                       & $0/100$                     & $0.131$                                  & $0/50$                         \\
            B1S2                              & \multicolumn{5}{c|}{---}              & \multicolumn{4}{c|}{$2.19$}                 & \multicolumn{3}{c}{---}                                                                                                                                                                                                                                                                                                                                                                          \\
            B1S3                              & $26$                                  & $27$                                        & $26$                                   & $26$                               & $19$                                       & $0.54$                             & $0.32$                             & $0.32$                             & $0.39$                                       & $36/100$                    & $0.666$                                  & $26/27$                        \\
            B1S4                              & $30$                                  & $30$                                        & $30$                                   & $30$                               & $4$                                        & \multicolumn{4}{c|}{$7.17$}        & $95/100$                           & $0.989$                            & $30/30$                                                                                                                                                \\
            \midrule
            B2S1                              & $31$                                  & $38$                                        & $46$                                   & $31$                               & $1$                                        & $0.77$                             & $1.25$                             & $1.33$                             & $1.12$                                       & $91/200$                    & $0.760$                                  & $31/50$                        \\
            B2S2                              & \multicolumn{5}{c|}{---}              & \multicolumn{4}{c|}{$1.71$}                 & \multicolumn{3}{c}{---}                                                                                                                                                                                                                                                                                                                                                                          \\
            B2S3                              & $45$                                  & $44$                                        & $45$                                   & $45$                               & $20$                                       & $0.40$                             & $1.45$                             & $0.27$                             & $0.70$                                       & $105/200$                   & $0.782$                                  & $44/45$                        \\
            B2S4                              & $49$                                  & $49$                                        & $49$                                   & $49$                               & $10$                                       & \multicolumn{4}{c|}{$7.11$}        & $193/200$                          & $0.988$                            & $49/49$                                                                                                                                                \\
            \midrule
            B3S1                              & $49$                                  & $55$                                        & $51$                                   & $49$                               & $0$                                        & $1.10$                             & $1.46$                             & $1.01$                             & $1.19$                                       & $208/300$                   & $0.868$                                  & $49/57$                        \\
            B3S2                              & \multicolumn{5}{c|}{---}              & \multicolumn{4}{c|}{$0.46$}                 & \multicolumn{3}{c}{---}                                                                                                                                                                                                                                                                                                                                                                          \\
            B3S3                              & $52$                                  & $52$                                        & $52$                                   & $52$                               & $3$                                        & $0.40$                             & $0.31$                             & $0.06$                             & $0.26$                                       & $228/300$                   & $0.897$                                  & $52/52$                        \\
            B3S4                              & $56$                                  & $56$                                        & $56$                                   & $56$                               & $14$                                       & \multicolumn{4}{c|}{$7.26$}        & $284/300$                          & $0.983$                            & $56/56$                                                                                                                                                \\
            \midrule
            B4S1                              & $57$                                  & $56$                                        & $56$                                   & $56$                               & $0$                                        & $1.96$                             & $2.35$                             & $1.65$                             & $1.99$                                       & $320/400$                   & $0.909$                                  & $56/57$                        \\
            B4S2                              & \multicolumn{5}{c|}{---}              & \multicolumn{4}{c|}{$1.42$}                 & \multicolumn{3}{c}{---}                                                                                                                                                                                                                                                                                                                                                                          \\
            B4S3                              & $56$                                  & $56$                                        & $56$                                   & $56$                               & $2$                                        & $0.24$                             & $0.19$                             & $0.15$                             & $0.19$                                       & $318/400$                   & $0.911$                                  & $56/56$                        \\
            B4S4                              & $56$                                  & $55$                                        & $56$                                   & $56$                               & $2$                                        & \multicolumn{4}{c|}{$5.65$}        & $381/400$                          & $0.985$                            & $55/56$                                                                                                                                                \\
            \bottomrule
        \end{tabular}
    }
\end{table}

\textbf{Time cost:}
The time costs of S2, S3, and S4 do not exhibit clear trends.
The time cost of S1 (Create/Edit Taxonomy) consistently increased through the batches, from an average of 0.73h at B1S1 to 1.99h at B4S1.
We interpret that the \textit{increased time cost results from the increased complexity of the taxonomy tree}.
From the coders' experience, the increased taxonomy tree complexity made it more laborious and required more discretion in deciding the image's taxa.
With the increased number of taxa, the division of the design space becomes finer, and the semantic distances between taxa decrease.

\textbf{Level of agreement:}
We examine the two metrics that reflect the level of agreement among coders on the image labels: exact match ratio ($Match$) and pairwise
averaged Jaccard score ($\overline{Jaccard}$).
We also examine the level of agreement on the taxonomy tree with the node intersection over union score ($Node~IoU$).

\begin{itemize}[leftmargin=*]
    \item \textbf{Source of disagreement}:
          \Cref{tab:coding-statistics} shows that at the beginning of each batch (B1S1, B2S1, B3S1, and B4S1), the number of mismatches is high.
          At B1S1, none of the images have the same label from all coders, and none of the nodes matches among coders.
          In addition to coders' disagreement on the image content, several factors contribute significantly to the mismatches:

          \begin{itemize}[leftmargin=*]
              \item \textbf{Inconsistent language use:}
                    Despite our effort to normalize the language use with a glossary (\iflabelexists{sec:glossary-of-terms}{\Cref{sec:glossary-of-terms}}{The supplemental material titled ``Glossary of Terms''}), we found discussions vital for coders to align the language use.
                    Minor discrepancies in the language use, such as plurality, can lead to mismatches.
                    In the first batch, the coders made different choices on using single or plural nouns to name a taxon.
                    Such discrepancies are easy to resolve but require a discussion.
                    Besides, the words that one uses to name a taxon may be infrequent in other coders' vocabulary.
                    This issue can be prominent when the coders use their second language to code.

              \item \textbf{Decision on when to split a taxon:}
                    A recurring source of systematic disagreements is deciding when to split a taxon.
                    For example, in the first batch, \coderIndexXinyuChen split bar chart into subcategories (e.g., \taxon{horizontal bar chart}), while \coderIndexWeiliZheng and \coderIndexYuZhang decided not to split the bar chart category as it contains less than 20 images.

              \item \textbf{Evaluation metrics:}
                    The evaluation metrics we use (number of mismatches, exact match ratio, and Jaccard score) rely on strict matching and do not consider the semantic similarity among labels.
          \end{itemize}

    \item \textbf{High but gradually decreasing disagreements at S1:}
          The discrepancy among the coders peaks at step 1 of each batch, as reflected in the exact match ratio and pairwise average Jaccard score.
          This is expected because at step 1 of each batch, a batch of images is added, and the coders code the images individually at this step.
          The coders may use different taxon names to describe the same design at step 1, and such discrepancies can only be resolved after discussions that take place at step 2.
          Meanwhile, the disagreement among the coders at step 1 gradually decreases throughout the batches, suggesting that the coders have \textit{converging opinions on the image labels}.
          Note that the number of mismatches at B4S1 is still relatively high.
          After adding 100 images at B4S1, the number of exact matches increased by only 36 (from 284 at B3S4 to 320 at B4S1).
          A major reason may be that the criteria for the exact match are strict.

    \item \textbf{Low but persistent disagreements at S4:}
          The discrepancy among the coders at step 4 of each batch is consistently low, as reflected in the exact match ratio and pairwise average Jaccard score.
          Additionally, the coders reached a consensus on the taxonomy tree structure at step 4 of each batch.
          Meanwhile, disagreement among the coders still exists at step 4 (i.e., the exact match ratio is not 100\%) despite the collective discussions for resolving conflicts.
          The persistent disagreements suggest that the coders have a small number of \textit{systematic disagreements on the image labels}.
          The disagreements reveal different perspectives of the coders, which is one of the purposes that we involve multiple coders (see \cref{sec:considerations}).
          \Cref{sec:on-the-coding-criteria} discusses some disagreements in more detail.
\end{itemize}

\subsection{On the Coding Criteria}
\label{sec:on-the-coding-criteria}

A major part of the coding involves deliberating on the criteria for defining the taxa, i.e., whether to include a taxon, the naming of the taxon, and its boundary that decides which visual designs should be included.
This section reflects on the findings and issues encountered during the coding process based on the coders' memos.
We mark whether the coders had consensus or dissensus on each issue.

\subsubsection{What Do We Mean When We Label (Consensus)}

The coders observed different perspectives on what it means to assign an image to a taxon in the following dimensions.

\textbf{Visual prominence:}
To assign an image to a taxon, should the visual design corresponding to the taxon be visually \textit{prominent} or \textit{present} in the image?
For example, for an image with multiple views, shall we assign the image to the taxon of the most prominent view (possibly in terms of the occupied area) or to the taxa of all views?
The coders agreed that an image should be assigned to all taxa present in the image.
The rationale is that the judgment on prominence can be subjective and hard in various cases.

\textbf{Design representativeness:}
To assign an image to a taxon, should the visual design corresponding to the taxon be \textit{representative} in the image?
A line chart can be seen as a special area chart with the area filled with the background color.
Shall we count a line chart as an area chart?
The coders agreed that for an image to be assigned to a taxon, the corresponding visual design needs to be representative.

\subsubsection{What is Considered Visualization (Dissensus)}

As described in \cref{sec:data-source}, the coders intend to exclude non-visualization images.
Throughout the coding process, there was consistent disagreement and debate among the coders on what is considered visualization.

\textbf{Controversy on the relation between maps and visualizations} was a key disagreement.
Should all maps be considered visualizations?
If not, what types of maps count as visualizations?
\coderIndexWeiliZheng argued that maps should be considered as visualizations only when employing multiple visual channels.
\coderIndexXinyuChen argued that all maps encode location information, and thus, all maps are visualizations.
\coderIndexYuZhang argued that maps are visualizations only when encoding abstract data beyond geographic information.
The views reflect narrow and broad definitions of visualization.

\textbf{Categorizations of maps in the literature} also reveal diverging views on whether maps are visualizations.
Pang et al.~\cite{Pang2016What} and Hogräfer et al.~\cite{Hograefer2020State} contrast map and visualization on a continuous spectrum, suggesting the two are distinct.
Friendly et al.~\cite{Friendly2010First} hold the view that maps with only cartographic representations of geographical features are not visualizations, while thematic maps that encode statistical data are visualizations.
\coderIndexYuZhang took this view.
Chen et al.~\cite{Chen2022Not,Chen2024Image} categorize maps into \taxon{generalized area representations} together with other visual representations with area encoding, such as area charts, implying that maps should be treated as visualizations.
\coderIndexWeiliZheng and \coderIndexXinyuChen took this view.

\textbf{Other discussions on the boundary of visualization} were also reflected in coders' memos.
Debates were raised on whether \taxon{table}, \taxon{timeline}, \taxon{bathymetric map}, and \taxon{star map} are visualizations.
The debates underscore the complexities in defining what constitutes a visualization, pin down characteristics of different visual representations, and inform the \iflabelexists{sec:taxon-definitions}{taxon definitions in~\Cref{sec:taxon-definitions}}{supplemental material titled ``Taxon Definitions''}.

\subsubsection{Should Historical Context Affect Labeling (Consensus)}

In the literature, the historical context is usually accounted for when judging whether an artifact is a visualization.
When tracing the history of visualization, the literature tends to hold looser criteria for the artifacts created earlier.
For example, the 10th-century graph showing planetary movement along the zodiac~\cite{unknown9011000Planetary} is sometimes regarded as the earliest known line chart.
By contemporary standards, it may not be considered a typical line chart, as it shows the physical movement rather than the trend of abstract data.
Through discussion, the coders agreed that the historical context should not be considered to reduce subjectivity and focus on the visual appearance of the image.

\subsubsection{How to Name a Taxon (Consensus)}

Throughout the coding process, coders gradually developed a consensus on the strategy for naming the taxa, as listed below.

\textbf{Avoid using a design dimension value as a taxon name:}
The coders discussed and agreed to name the taxon corresponding to bar charts as ``bar chart'' instead of ``bar''.
The consideration is that ``bar'' refers to a visual mark and is a value of a design dimension.
If the taxon names are all design dimensions, the number of taxa may grow exponentially because to depict different designs, each dimension may need to take different values.
The large number of taxa would make the taxonomy less manageable.
If only part of the taxon names are design dimensions, the naming would be inconsistent.

\textbf{Avoid using a data structure as a taxon name:}
The coders discussed and agreed to name the taxon corresponding to node-link diagrams as ``node-link diagram'' instead of ``graph''.
The consideration is that ``graph'' refers to a data structure.
There is a limited number of data structure names.
Thus, consistently using data structures as taxon names would limit the number of taxa.
If only part of the taxon names are data structures, the naming would be inconsistent.

    \section{Discussion}

This section discusses considerations, limitations, and future directions.

\textbf{On the reusability of the protocol and \systemName:}
Our protocol and \systemName are designed to be generic and reusable for other image taxonomy development scenarios.
One may use them for visualization taxonomies in dimensions other than visual design, such as the theme.

\textbf{On the machine assistance:}
\Cref{sec:machine-assistance} describes two types of machine assistance, clustering and semantic cluster overview.
One may consider other potential types of machine assistance, e.g., integrate the prediction method in \cref{sec:predicting-taxonomy-labels} into \systemName.
Meanwhile, machine assistance comes with not only benefits but also potential issues.

\begin{itemize}[leftmargin=*]
    \item \textbf{Advantage -- efficiency:}
          Delegating specific tasks to machines that were originally conducted manually may improve the efficiency of coders.
          In data labeling and coding, providing default labels is a common type of machine assistance~\cite{Zhang2021MI3,Zhang2022OneLabeler,Zhang2024Simulation}.

    \item \textbf{Disadvantage -- anchoring effect:}
          Machine assistance may introduce over-reliance~\cite{Goddard2012Automation}.
          In data labeling and coding, machine-predicted default labels may cause an anchoring effect~\cite{Tversky1974Judgment} and bias the annotator's decision, even for expert annotators~\cite{Levy2021Assessing}.
          Anchoring effect may reduce the quality and variation of the coders' annotations.
          Thus, we intentionally avoid the attempt to give predictions in \systemName to avoid the anchoring effect.
\end{itemize}

\textbf{Extending the comparison of visualization taxonomies:}
We plan to extend our comparison with existing schemes on high-level characteristics (\cref{sec:characterizing-schemes}) and content (\cref{sec:distribution-analysis} and \cref{sec:comparison-with-existing-schemes}).
An ample scope is to increase the coverage of the literature.
A systematic comparison may involve aligning the language use in different taxonomies and attributing the cause of their similarities and differences.

\textbf{Taxonomy for extending and improving dataset:}
The taxonomy may extend and improve the quality of the coded visualization dataset in the following aspects.

\begin{itemize}[leftmargin=*]
    \item \textbf{Serve as search keywords:}
          The taxonomy labels may be integrated into the OldVisOnline dataset~\cite{Zhang2024OldVisOnline} and serve as search keywords.

    \item \textbf{Identify non-visualizations:}
          The coding process helped identify a small portion of non-visualization images, which may be excluded from the next version of the OldVisOnline dataset.

    \item \textbf{Clarify scope:}
          Our taxonomy serves as an overview of the OldVisOnline dataset.
          The taxonomy establishes boundaries between visualization and non-visualization, and among different chart types.
          The boundaries clarify the inclusions and scope of the dataset.
          In this sense, the taxonomy serves as an enumerative definition of historical visualization in the OldVisOnline dataset.
\end{itemize}

\textbf{Limitation of the corpus:}
As mentioned in \cref{sec:data-source}, the corpus we used for coding focuses on imagery artifacts hosted in Europe and the United States.
Future work is necessary to synthesize a corpus of visualizations from other regions and cultures to extend our taxonomy.

\textbf{Toward a comprehensive taxonomy:}
We do not see our taxonomy as completed but rather as a starting point.
A more comprehensive taxonomy can be made possible by scaling up our method to code larger corpora in the future.
Future coders may reuse our coding method (\cref{sec:coding-method}) and the \systemName system (\cref{sec:system}), and extend our taxonomy (\cref{sec:the-final-taxonomy}) by looking for outliers that do not fit.
Specifically, one may conduct such extensions by using our taxonomy as:
\begin{itemize}[leftmargin=*]
    \item the initial taxonomy in the coding process, or
    \item an external information source to inform the coding practice, which resembles our use of existing taxonomies in the glossary (\cref{sec:considerations}).
\end{itemize}

\textbf{Alternative scheme to structure design space:}
This work uses taxonomy to structure the design space of historical visualizations.
Other schemes, such as a grammar~\cite{Ying2024VAID}, may also serve the purpose.
Different schemes may have different tradeoffs in aspects such as efficiency of development, generalizability, and supported usage scenarios.

\textbf{Alternative protocols and methodologies to develop the scheme:}
Our protocol (\cref{sec:coding-method}) is designed to be lightweight and focuses on developing a descriptive (instead of explanatory) framework.
If one is to develop an explanatory framework (e.g., focusing on explaining why a visual design emerged given its historical context), one may explore protocols and methodologies alternative to ours (e.g., grounded theory).

    \section{Conclusion}

To understand \textit{what visual designs have been used in historical visualizations}, we resort to taxonomy.
As a step toward a comprehensive taxonomy of historical visualizations, we describe an empirical method for developing image taxonomy.
Our method involves a coding protocol and the \systemName system for image taxonomy labeling and comparison.
Our method is designed to be generic and reusable.
The protocol and \systemName system support progressive updates of the corpus and the taxonomy.
We use our protocol and the \systemName system to develop a taxonomy from \nOldVisLabeled historical visualizations in the OldVisOnline dataset.
We reflect on the coding process and observe that the taxonomy and the opinions among coders exhibit a converging trend.
Our current taxonomy is limited to our corpus.
Future research may reuse our protocol and taxonomy labeling tool to extend our taxonomy.

    \section*{Supplemental Material}

\iflabelexists{sec:glossary-of-terms}{\Cref{sec:glossary-of-terms}}{The supplemental material titled ``Glossary of Terms''}
describes a glossary of visualization terms the coders consulted in the coding process described in \cref{sec:coding-method}.
\iflabelexists{sec:taxon-definitions}{\Cref{sec:taxon-definitions}}{The supplemental material titled ``Taxon Definitions''}
describes the definition of the taxons the coders derived in the coding process described in \cref{sec:coding-method}.

    \acknowledgments{
    This work is supported by NSFC No. 62272012, No.62472099, and No.62202105.
    It is also partially supported by Wuhan East Lake High-Tech Development Zone National Comprehensive Experimental Base for Governance of Intelligent Society.
}

    \bibliographystyle{abbrv-doi-hyperref}

    \bibliography{assets/bibs/papers,assets/bibs/historical-visualizations}
\fi

\ifx\hideappendix\undefined

    \pagebreak
    \appendix %
    \section{Glossary of Terms}
\label{sec:glossary-of-terms}

\Cref{tab:glossary} describes a glossary of visualization terms used in related works.
The glossary is used as a reference in the coding process to improve the consistency of language use by different coders.

\begin{table*}[ht]
\caption{
\textbf{Glossary:}
The terms are extracted from \nGlossaryPaper papers related to visualization classification schemes.
The ``Terms Extracted From'' column describes the location of the terms in the original papers.
The glossary is used for coders' reference in the coding process.
}
\label{tab:glossary}
\centering
\scriptsize
\renewcommand{\arraystretch}{1} %
\begin{tabular}{c c c c c p{0.55\linewidth}}
\toprule
\textbf{ID} & \textbf{Paper} & \textbf{Venue} & \textbf{Year} & \textbf{Terms Extracted From} \\ 
\toprule
1 & Lohse1994Classification~\cite{Lohse1994Classification} & CACM & 1994 & Discussion \\ 
\midrule
\multicolumn{5}{p{0.9\linewidth}}{\raggedright\textbf{Terms:} ``graph'', ``table'', ``numerical table'', ``graphical table'', ``time chart'', ``network chart'', ``diagram'', ``structure diagram'', ``process diagram'', ``map'', ``cartogram'', ``icon'', ``photo-realistic picture''} \\ 
\toprule
2 & Chi2000Taxonomy~\cite{Chi2000Taxonomy} & VIS & 2000 & Section 4 \\ 
\midrule
\multicolumn{5}{p{0.9\linewidth}}{\raggedright\textbf{Terms:} ``scientific visualization'', ``geographical-based info visualization'', ``2D'', ``multi-dimensional plot'', ``information landscape and space'', ``tree'', ``network'', ``text'', ``web visualization'', ``visualization spreadsheet''} \\ 
\toprule
3 & Fujishiro2000GADGET/IV~\cite{Fujishiro2000GADGET_IV} & InfoVis & 2000 & Figure 1 \\ 
\midrule
\multicolumn{5}{p{0.9\linewidth}}{\raggedright\textbf{Terms:} ``bar chart'', ``city space'', ``stacked bar chart'', ``polar bar chart'', ``tile bars'', ``error bars'', ``city scape'', ``histogram'', ``line chart'', ``stepped surface chart'', ``line chart'', ``link'', ``link net'', ``parallel coordinate'', ``pie chart'', ``scatter plot'', ``scatter cube'', ``tile graph'', ``tree'', ``cone tree'', ``treemap'', ``venn diagram'', ``topography'', ``thumbnail'', ``timetable'', ``datasheet'', ``distortion'', ``geographical map''} \\ 
\toprule
4 & Heer2010Tour~\cite{Heer2010Tour} & CACM & 2010 & all sections \\ 
\midrule
\multicolumn{5}{p{0.9\linewidth}}{\raggedright\textbf{Terms:} ``time-series data'', ``index chart'', ``stacked graph'', ``small multiples'', ``horizon graph'', ``statistical distribution'', ``histogram'', ``box-and-whisker plot'', ``stem-and-leaf plot'', ``Q-Q plot'', ``SPLOM (scatter plot matrix)'', ``parallel coordinates'', ``map'', ``flow map'', ``choropleth map'', ``graduated symbol map'', ``cartogram'', ``hierarchy'', ``node-link diagram'', ``adjacency diagram'', ``enclosure diagram'', ``network'', ``force-directed layout'', ``arc diagram'', ``matrix view''} \\ 
\toprule
5 & Borkin2013What~\cite{Borkin2013What} & TVCG & 2013 & Table 1 \\ 
\midrule
\multicolumn{5}{p{0.9\linewidth}}{\raggedright\textbf{Terms:} ``area'', ``area chart'', ``overlap area chart'', ``stacked area chart'', ``proportional area chart'', ``aligned area chart'', ``centered area chart'', ``overlapped area chart'', ``stacked and linked area chart'', ``bar'', ``bar chart'', ``grouped bar chart'', ``stacked bar chart'', ``circular bar chart'', ``waterfall chart'', ``bullet graph'', ``circle'', ``belt chart'', ``donut chart'', ``pie chart'', ``sector graph'', ``diagram'', ``flow chart'', ``illustration or rendering'', ``Sankey diagram'', ``timeline'', ``Venn diagram'', ``distribution'', ``box-and-whisker plot'', ``distribution curve'', ``dot array'', ``histogram'', ``point graph'', ``stem-and-leaf plot'', ``stripe graph'', ``tally graph'', ``grid \& matrix'', ``heatmap'', ``line'', ``contour graph'', ``density graph'', ``line graph'', ``circular line graph'', ``trend line (and residual graph)'', ``slopegraph'', ``star plot'', ``surface graph'', ``vector graph'', ``map'', ``flow map'', ``geographic map'', ``street map'', ``statistical map'', ``choropleth map'', ``contour map'', ``distorted map'', ``plotted map'', ``point'', ``dot plot'', ``scatter plot'', ``bubble chart'', ``(trend line and) residual graph'', ``trilinear scatter plot'', ``table'', ``text chart'', ``text'', ``phrase net'', ``word cloud'', ``word tree'', ``tree \& network'', ``graph'', ``matrix representation'', ``tree'', ``treemap'', ``hive graph'', ``hierarchical edge bundling''} \\ 
\toprule
6 & Battle2018Beagle~\cite{Battle2018Beagle} & CHI & 2018 & Table 1 \\ 
\midrule
\multicolumn{5}{p{0.9\linewidth}}{\raggedright\textbf{Terms:} ``area'', ``bar'', ``box'', ``bubble'', ``chord'', ``donut'', ``heatmap'', ``geographic map'', ``graph'', ``hexabin'', ``line'', ``radial'', ``pie'', ``sankey'', ``scatter'', ``treemap'', ``voronoi'', ``waffle'', ``word cloud'', ``sunburst'', ``stream graph'', ``parallel coordinates'', ``contour'', ``filled-line''} \\ 
\toprule
7 & Crisan2018systematic~\cite{Crisan2018systematic} & Bioinformatics & 2018 & Figure 4 \\ 
\midrule
\multicolumn{5}{p{0.9\linewidth}}{\raggedright\textbf{Terms:} ``common statistical chart'', ``bar chart'', ``standard bar chart'', ``stacked bar chart'', ``divergent bar chart'', ``epidemic curve'', ``diversity chart'', ``LefSe plot'', ``line chart'', ``bootscan'', ``Kaplan-Meier'', ``skyline plot'', ``scatter plot'', ``root-to-tip'', ``ordination plot'', ``Q-Q plot'', ``distribution plot'', ``histogram'', ``PDF'', ``boxplot'', ``swarm plot'', ``pie chart'', ``Venn diagram'', ``colour chart'', ``category stripe'', ``heatmap'', ``density plot'', ``relational chart'', ``node-link'', ``eBurst'', ``social network'', ``molecular network'', ``minimum spanning tree'', ``flow diagram'', ``chord diagram'', ``Sankey diagram'', ``temporal chart'', ``streamgraph'', ``absolute streamgraph'', ``relative streamgraph'', ``timeline'', ``spatial chart'', ``geographic map'', ``choropleth map'', ``interior map'', ``tree chart'', ``phylogenetic tree'', ``rooted phylogenetic tree'', ``unrooted phylogenetic tree'', ``dendrogram'', ``clonal tree'', ``genomic chart'', ``genomic map'', ``linear genomic map'', ``radial genomic map'', ``alignment'', ``composition plot'', ``sequence logo plot'', ``other chart'', ``table'', ``image'', ``gel image'', ``general image'', ``miscellany''} \\ 
\toprule
8 & Deng2023VisImages~\cite{Deng2023VisImages} & TVCG & 2023 & Table 2 \\ 
\midrule
\multicolumn{5}{p{0.9\linewidth}}{\raggedright\textbf{Terms:} ``area'', ``area chart'', ``proportional area chart (PAC)'', ``bar'', ``bar chart'', ``circle'', ``donut chart'', ``pie chart'', ``diagram'', ``flow diagram'', ``chord diagram'', ``Sankey diagram'', ``Venn diagram'', ``statistic'', ``box plot'', ``error bar'', ``stripe graph'', ``table'', ``line'', ``contour graph'', ``line chart'', ``storyline'', ``polar plot'', ``parallel coordinate (PCP)'', ``surface graph'', ``vector graph'', ``map'', ``point'', ``scatter plot'', ``grid \& matrix'', ``heatmap'', ``matrix'', ``text'', ``phrase net'', ``word cloud'', ``word tree'', ``graph \& tree'', ``graph'', ``tree'', ``treemap'', ``hierarchical edge bundling (HEB)'', ``sunburst/icicle plot'', ``special'', ``glyph-based visualization'', ``unit visualization''} \\ 
\toprule
9 & Chen2024Image~\cite{Chen2024Image} & arXiv & 2024 & Table 1 \\ 
\midrule
\multicolumn{5}{p{0.9\linewidth}}{\raggedright\textbf{Terms:} ``generalized bar representation'', ``point-based representation'', ``line-based representation'', ``node-link tree/graph'', ``network'', ``mesh'', ``generalized area representation'', ``surface-based representation and volume'', ``generalized matrix/grid'', ``continuous color and grey-scale representation'', ``glyph-based representation'', ``text-based representation''} \\ 
\toprule
10 & Arunkumar2024Image~\cite{Arunkumar2024Image} & VIS & 2024 & Figure 3 \\ 
\midrule
\multicolumn{5}{p{0.9\linewidth}}{\raggedright\textbf{Terms:} ``bar'', ``stacked bar'', ``radial bar'', ``mosaic'', ``grouped bar'', ``map'', ``plotted map'', ``choropleth'', ``geographical'', ``contour'', ``area'', ``proportional area'', ``stacked area'', ``line'', ``slope'', ``contour'', ``parallel coordinates'', ``point'', ``scatter'', ``dot'', ``bubble'', ``circle'', ``pie'', ``belt'', ``sector'', ``donut'', ``chord'', ``spider'', ``tree'', ``treemap'', ``circle packing'', ``grid'', ``heatmap'', ``distribution'', ``histogram'', ``box'', ``stripe'', ``beeswarm'', ``curve'', ``network'', ``hierarchical edge bundling'', ``graph'', ``table'', ``diagram'', ``isotype'', ``illustration'', ``alluvial/Sankey'', ``timeline'', ``Chernoff'', ``Venn''} \\ 
\toprule
\end{tabular}
\label{tab:mytable1}
\end{table*}

    \section{Taxon Definitions}
\label{sec:taxon-definitions}

\Cref{fig:examples} shows example historical images belonging to each leaf taxon in the taxonomy tree.
\Cref{tab:taxon-definitions} provides definitions for the taxa.
Note that the definitions are not meant to be conclusive, but rather as a scaffold for coders to pin down the characteristics of each taxon.
We expect the definitions to evolve through future research.

\begin{figure*}[!htbp]
    \centering
    \includegraphics[width=0.9\linewidth]{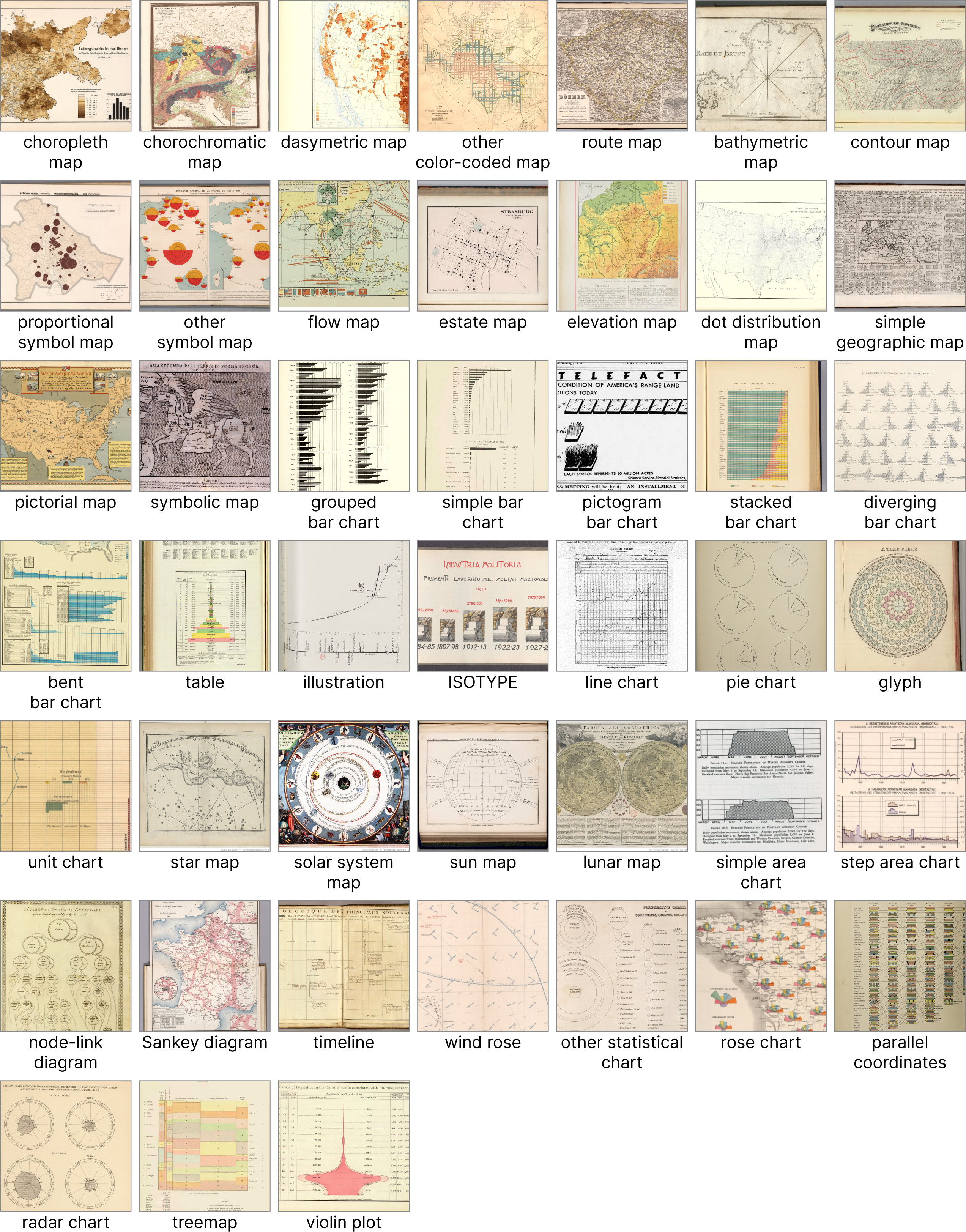}
    \caption{
        \textbf{Example images in each leaf taxon:}
        Among the \nTaxaFinalAllVis taxa (excluding \taxon{root}, \taxon{non-visualization}, and the subcategories of \taxon{non-visualization}), \nTaxaFinalLeafVis are leaf nodes in the taxonomy tree.
        We show an example image of each leaf node in this image matrix.
        Note that some images may have been assigned to more than one leaf node by the coders.
        Some images are cropped to show representative design of specific taxon more clearly.
        The definition of each leaf taxon can be found in \cref{tab:taxon-definitions}.
    }
    \label{fig:examples}
\end{figure*}

\newlength\lw
\setlength\lw{\dimexpr .35\textwidth - 2\tabcolsep}
\newlength\rw
\setlength\rw{\dimexpr .65\textwidth - 2\tabcolsep}

\begin{table*}[!htbp]
    \caption{
    \textbf{Taxon definitions:}
    This table provides the definition of each taxon.
    Most definitions are structured as: \inlinecode{A/An [``visual representation'' or a category] that (``commonly'') uses [what visual design] to represent [what data (in terms of theme or other properties)]. ([additional characteristics or comments].)}
    where ``[]'' denotes variable content and ``()'' denotes optional.
    We use ``commonly'' to refer to a characteristic that is not definitive.
    Some definitions are structured as: \inlinecode{A/An [``visual representation'' or a category] that [has what characteristics]. ([additional characteristics or comments].)}
    We use \textbf{bold text} to highlight leaf nodes in the taxonomy tree.
    For each leaf node (excluding subcategories of \taxon{non-visualization}), we give a link to an example image that we coded.
    The titles of the example images are extracted from the image sources.
    }
    \label{tab:taxon-definitions}
    \centering
    \scriptsize
    \begin{tabular}{p{\lw}p{\rw}}
        \toprule
        \textbf{Taxon}                                            & \textbf{Definition}                                                                                                                                                                                                                                                                                                                                                                                                                           \\
        \midrule
        map                                                       & A [visual representation] that uses [spatially distributed graphical elements] to represent [geographical data, including regions, locations, and their associated non-spatial data, such as demographics and networks].                                                                                                                                                                                                                      \\
        \midrule
        map > color-coded map                                     & A [map] that uses [colored regions] to represent [data with both spatial and non-spatial aspects, such as a spatial distribution of population density].                                                                                                                                                                                                                                                                                      \\
        \midrule
        map > color-coded map > \leafTaxon{choropleth map}        & A [map] that uses [uniformly colored, shaded, or patterned regions that adhere to pre-defined boundaries, such as national boundaries,] to represent [quantitative data associated with these regions]. \href{https://www.davidrumsey.com/luna/servlet/detail/RUMSEY~8~1~342996~90111071}{Example: ``Leberegelseuche bei den Rindern auf Grund der Feststellungen der Schlachtvieh- und Fleischbeschau im Jahre 1928''}.                      \\
        \midrule
        map > color-coded map > \leafTaxon{chorochromatic map}    & A [map] that uses [uniformly colored, shaded, or patterned regions that may not adhere to pre-defined boundaries] to represent [categorical or ordinal data associated with these regions]. \href{https://www.davidrumsey.com/luna/servlet/detail/RUMSEY~8~1~294598~90065469}{Example: ``No. 12. Deutschland ... Geognostische Ubersichtskarte''}.                                                                                            \\
        \midrule
        map > color-coded map > \leafTaxon{dasymetric map}        & A [map] that uses [uniformly colored, shaded, or patterned regions that may not adhere to pre-defined boundaries] to represent [quantitative data associated with these regions]. \href{https://archive.org/details/statisticalatlas00unitrich/page/n241}{Example: ``Statistical Atlas of the United States 1900 - Plate No. 80''}.                                                                                                           \\
        \midrule
        map > color-coded map > \leafTaxon{other color-coded map} & A [map] that [matches the definition of ``color-coded map'' while mismatches the definition of the other subcategories of ``color-coded map'']. \href{https://www.loc.gov/item/87695629/}{Example: ``Map of the city of Washington''}.                                                                                                                                                                                                        \\
        \midrule
        map > \leafTaxon{route map}                               & A [map] that commonly uses [lines and symbols that prioritize clarity over geographic accuracy] to represent [transportation routes and networks, such as roads and railways, focusing on human-made rather than natural features]. \href{https://www.davidrumsey.com/luna/servlet/detail/RUMSEY~8~1~319733~90088643}{Example: ``Konigreich Bohmen''}.                                                                                        \\
        \midrule
        map > \leafTaxon{bathymetric map}                         & A [map] that commonly uses [numeric labels, contour lines, or color gradients] to represent [underwater depth]. \href{https://www.davidrumsey.com/luna/servlet/detail/RUMSEY~8~1~284312~90056818}{Example: ``Pl. 22. Brusc, France''}.                                                                                                                                                                                                        \\
        \midrule
        map > \leafTaxon{contour map}                             & A [map] that uses [contour lines] to represent [points of equal value in a scalar field, such as elevation and temperature, in the physical space]. \href{https://www.davidrumsey.com/luna/servlet/detail/RUMSEY~8~1~26389~1100060}{Example: ``Penn. climatological map''}.                                                                                                                                                                   \\
        \midrule
        map > symbol map                                          & A [map] that uses [symbols, such as shapes, icons, and glyphs,] to represent [non-spatial data associated with locations or regions].                                                                                                                                                                                                                                                                                                         \\
        \midrule
        map > symbol map > \leafTaxon{proportional symbol map}    & A [map] that uses [symbols of varying sizes] to represent [quantitative data associated with locations or regions]. [The size of each symbol is proportional to the data value]. \href{https://www.davidrumsey.com/luna/servlet/detail/RUMSEY~8~1~329643~90098050}{Example: ``Gyáripari telepek. (Folytatás.) - Fabrikindustrieanlagen. (Fortsetzung.) 1932''}.                                                                               \\
        \midrule
        map > symbol map > \leafTaxon{other symbol map}           & A [map] that [matches the definition of ``symbol map'' while mismatches the definition of the other subcategory of ``symbol map'']. \href{https://www.davidrumsey.com/luna/servlet/detail/RUMSEY~8~1~313766~90082656}{Example: ``Commerce special de la France de 1891 a 1896''}.                                                                                                                                                             \\
        \midrule
        map > \leafTaxon{flow map}                                & A [map] that uses [lines or arrows, sometimes with varying widths, that prioritize clarity over geographic accuracy] to represent [spatial movement of entities, such as people and goods]. \href{https://gallica.bnf.fr/ark:/12148/bpt6k1100178q/f241.item}{Example: ``Essai d'Atlas statistique de l'Indochine française : Indochine physique, population, administration, finances, agriculture, commerce, industrie - Page 246 to 247''}. \\
        \midrule
        map > \leafTaxon{estate map}                              & A [map] that commonly uses [boundary lines] to represent [land parcels]. \href{https://www.davidrumsey.com/luna/servlet/detail/RUMSEY~8~1~207927~3003519}{Example: ``Strasburg, Virginia''}.                                                                                                                                                                                                                                                  \\
        \midrule
        map > \leafTaxon{elevation map}                           & A [map] that commonly uses [numeric labels, contour lines, or color gradients] to represent [elevations above a reference level]. \href{https://gallica.bnf.fr/ark:/12148/bpt6k6571542q/f70.item}{Example: ``France physique - cours d'eau''}                                                                                                                                                                                                 \\
        \midrule
        map > \leafTaxon{dot distribution map}                    & A [map] that uses [dots of uniform size] to represent [the distribution and density of entities, such as population]. [Each dot typically represents a fixed quantity of entities]. \href{https://archive.org/details/statisticalatlas00sloa/page/348}{Example: ``Statistical Atlas of the United States 1914 - Plate No. 348''}.                                                                                                             \\
        \midrule
        map > \leafTaxon{simple geographic map}                   & A [map] that [does not represent non-spatial data]. \href{https://www.davidrumsey.com/luna/servlet/detail/RUMSEY~8~1~321107~90090383}{Example: ``Tom II. No. 1. Carte de l'etendue de l'Empire Romain''}.                                                                                                                                                                                                                                     \\
        \midrule
        map > \leafTaxon{pictorial map}                           & A [map] that uses [pictorial illustrations and artistic styles] to represent [geographical data and associated non-spatial data]. \href{https://www.davidrumsey.com/luna/servlet/detail/RUMSEY~8~1~264737~5524566}{Example: ``Map of America's making''}.                                                                                                                                                                                     \\
        \midrule
        map > \leafTaxon{symbolic map}                            & A [map] that uses [metaphoric symbols] to represent [geographical data and associated non-spatial data].  \href{https://imagesonline.bl.uk/asset/162414}{Example: ``Asia secunda pars terrae in forma Pegasir''}.                                                                                                                                                                                                                             \\
        \midrule
    \end{tabular}
\end{table*}

\begin{table*}[!htbp]
    \ContinuedFloat %
    \caption{continued}
    \centering
    \scriptsize
    \begin{tabular}{p{\lw}p{\rw}}
        \toprule
        \textbf{Taxon}                                     & \textbf{Definition}                                                                                                                                                                                                                                                                                                                                                                                                                                       \\
        \midrule
        bar chart                                          & A [visual representation] that uses [rectangular bars or silhouettes with varying lengths or heights] to represent [quantitative data across categories or intervals]. [This definition expands the common notion of ``bar chart'' to include histogram].                                                                                                                                                                                                 \\
        \midrule
        bar chart > \leafTaxon{grouped bar chart}          & A [bar chart] that uses [clusters of grouped bars] to represent [quantitative data across categories and subcategories]. [Each cluster corresponds to a category, and each bar in a cluster corresponds to a subcategory]. \href{https://archive.org/details/statisticalatlas00sloa_0/page/171}{Example: ``Statistical Atlas of the United States 1924 - Plate No. 142''}.                                                                                \\
        \midrule
        bar chart > \leafTaxon{simple bar chart}           & A [bar chart] that uses [rectangular bars with varying lengths or heights] to represent [the data values of a quantitative variable across categories or intervals]. [No stacking or grouping is applied to the bars]. \href{https://archive.org/details/CAT10508326/page/n63}{Example: ``Forest service atlas. Extracts from the statistical volume of the Forest Atlas for the fiscal year 1907 - Page 27''}.                                           \\
        \midrule
        bar chart > \leafTaxon{pictogram bar chart}        & A [bar chart] that uses [rectangular silhouettes of pictograms with varying lengths or heights] to represent [quantitative data across categories or intervals].
        \href{https://modley-telefact-1939-1945.tumblr.com/post/614489540077486080/rudolf-modley-pictograph}{Example: ``Telefact - The Conditions of America's Range Land Conditions Today''}.                                                                                                                                                                                                                                                                                                                         \\
        \midrule
        bar chart > \leafTaxon{stacked bar chart}          & A [bar chart] that uses [bars divided into stacked segments] to represent [the composition of quantitative data across categories or intervals]. \href{https://www.davidrumsey.com/luna/servlet/detail/RUMSEY~8~1~32196~1151538}{Example: ``Farm area by tenure''}.                                                                                                                                                                                       \\
        \midrule
        bar chart > \leafTaxon{diverging bar chart}        & A [bar chart] that uses [pairs of bars extending in opposite directions from a baseline] to represent [the data values of two quantitative variables across categories]. \href{https://archive.org/details/statisticalatla00unit/page/n151}{Example: ``Statistical atlas of the United States based on the results of the ninth census 1870 with contributions from many eminent men of science and several departments of the government - PI. XXXIX''}. \\
        \midrule
        bar chart > \leafTaxon{bent bar chart}             & A [bar chart] that uses [bars and bent bars] to represent [quantitative data across categories or intervals]. [The bars are bent to fit in limited space]. \href{https://www.davidrumsey.com/luna/servlet/detail/RUMSEY~8~1~32791~1152169}{Example: ``Gold mining regions''}.                                                                                                                                                                             \\
        \midrule
        \leafTaxon{table}                                  & A [visual representation] that uses [rows and columns of cells] to represent [structured data]. [Typically, more than two rows and columns are used]. \href{https://www.davidrumsey.com/luna/servlet/detail/RUMSEY~8~1~20795~560076}{Example: ``U.S. wool manufacture, 1890, 1880''}.                                                                                                                                                                     \\
        \midrule
        \leafTaxon{illustration}                           & A [visual representation] that commonly uses [drawings, sketches, or paintings] to represent [visual explanations of concepts, plans, processes, or scenes]. \href{https://gallica.bnf.fr/ark:/12148/bpt6k324466r/f176.item}{Example: ``Atlas Hydroélectrique Profil no. 7''}.                                                                                                                                                                            \\
        \midrule
        non-visualization                                  & An [image] that [does not qualify as a visualization]. [This category is used to mark images that fall outside the scope of a visualization taxonomy].                                                                                                                                                                                                                                                                                                    \\
        \midrule
        non-visualization > \leafTaxon{plain illustration} & A [non-visualization] that [falls in the ``illustration'' category].                                                                                                                                                                                                                                                                                                                                                                                      \\
        \midrule
        non-visualization > \leafTaxon{plain text}         & A [non-visualization] that [is a text page].                                                                                                                                                                                                                                                                                                                                                                                                              \\
        \midrule
        non-visualization > \leafTaxon{plain map}          & A [non-visualization] that [falls in the ``map'' category].                                                                                                                                                                                                                                                                                                                                                                                               \\
        \midrule
        non-visualization > \leafTaxon{plain table}        & A [non-visualization] that [falls in the ``table'' category].                                                                                                                                                                                                                                                                                                                                                                                             \\
        \midrule
        \leafTaxon{ISOTYPE}                                & A [visual representation] that uses [pictorial symbols with varying sizes or quantities] to represent [quantitative data]. \href{https://www.davidrumsey.com/luna/servlet/detail/RUMSEY~8~1~347143~90114621}{Example: ``Industria molitoria: Frumento lavorato nei molini nazionali (ql)''}.                                                                                                                                                              \\
        \midrule
        \leafTaxon{line chart}                             & A [visual representation] that uses [a series of connected straight or curved line segments] to represent [quantitative data across a continuous scale]. \href{https://archive.org/details/NavalMedicalBulletin471947/page/n670}{Example: ``Clinical Chart''}.                                                                                                                                                                                            \\
        \midrule
        \leafTaxon{pie chart}                              & A [visual representation] that uses [a circle, a circular sector, or a donut divided into slices] to represent [proportions of categories]. [The angle of each slice is proportional to the corresponding proportion]. \href{https://archive.org/details/b31366661/page/n334}{Example: ``Statistical Atlas of the United States 1900 - Plate No. 127''}.                                                                                                  \\
        \midrule
        \leafTaxon{glyph}                                  & A [visual representation] that uses [symbols with more than two variable visual attributes, such as size, color, and orientation,] to represent [multi-dimensional data]. \href{https://www.davidrumsey.com/luna/servlet/detail/RUMSEY~8~1~37203~1210183}{Example: ``Time table''}.                                                                                                                                                                       \\
        \midrule
        \leafTaxon{unit chart}                             & A [visual representation] that uses [symbols with varying quantities] to represent [quantitative data]. [Each symbol typically represents a fixed quantity]. \href{https://www.davidrumsey.com/luna/servlet/detail/RUMSEY~8~1~315690~90084491}{Example: ``1842-1843: Bildliche Darstellung der Geschichte der ausschl. privil. Kaiser Ferdinand's Nordbahn''}.                                                                                            \\
        \midrule
        celestial map                                      & A [visual representation] that commonly uses [planetary orbits and pictorial symbols] to represent [positions, movements, states, and relations of celestial bodies].                                                                                                                                                                                                                                                                                     \\
        \midrule
        celestial map > \leafTaxon{star map}               & A [celestial map] that commonly uses [dot symbols] to represent [positions of stars and shapes of constellations]. \href{https://gallica.bnf.fr/ark:/12148/btv1b53200892d/f256.item}{Example: ``Atlas caelestis containing the systems and theories of the planets, the constellations of the starrs and other phenomena's of the heavens with nessesary [sic] tables relating there to collected''}.                                                     \\
        \midrule
        celestial map > \leafTaxon{solar system map}       & A [celestial map] that commonly uses [planetary orbits and pictorial symbols] to represent [positions, movements, states, and relations of celestial bodies in the solar system]. \href{https://gallica.bnf.fr/ark:/12148/bpt6k53406427/f10.item}{Example: ``Harmonia macrocosmica sev atlas universalis et novus''}.                                                                                                                                     \\
        \midrule
    \end{tabular}
\end{table*}

\begin{table*}[!htbp]
    \ContinuedFloat %
    \caption{continued}
    \centering
    \scriptsize
    \begin{tabular}{p{\lw}p{\rw}}
        \toprule
        \textbf{Taxon}                             & \textbf{Definition}                                                                                                                                                                                                                                                                                                                                                                                                                             \\
        \midrule
        celestial map > \leafTaxon{sun map}        & A [celestial map] that commonly uses [detailed imageries and symbols] to represent [surface characteristics of the sun, such as sunspots]. \href{https://www.davidrumsey.com/luna/servlet/detail/RUMSEY~8~1~286088~90058605}{Example: ``Chart for Sun Spot Observations No. 3''}.                                                                                                                                                               \\
        \midrule
        celestial map > \leafTaxon{lunar map}      & A [celestial map] that commonly uses [detailed imageries and pictorial symbols] to represent [features and states of the moon, such as surface characteristics, phases, and orbital positions]. \href{https://gallica.bnf.fr/ark:/12148/btv1b85922715/f1.item}{Example: ``Atlas coelestis. Tabula selenographica''}.                                                                                                                            \\
        \midrule
        area chart                                 & A [visual representation] that uses [filled areas under a series of connected straight or curved line segments] to represent [quantitative data across a continuous scale].                                                                                                                                                                                                                                                                     \\
        \midrule
        area chart > \leafTaxon{simple area chart} & An [area chart] that [mismatches the definition of the other subcategory of ``area chart'']. \href{https://www.davidrumsey.com/luna/servlet/detail/RUMSEY~8~1~347250~90114723}{Example: ``Evacuee population of Fresno Assembly Center''}.                                                                                                                                                                                                      \\
        \midrule
        area chart > \leafTaxon{step area chart}   & An [area chart] that uses [filled areas under a stepped line with changes at discrete intervals] to represent [quantitative data across a continuous scale]. \href{https://www.davidrumsey.com/luna/servlet/detail/RUMSEY~8~1~331328~90099759}{Example: ``Fertözö betegségek: hasi hagymáz. - Infektiöse Krankheiten: Typhus Abdominalis.''}                                                                                                    \\
        \midrule
        \leafTaxon{node-link diagram}              & A [visual representation] that uses [nodes and links connecting nodes] to represent [a network of entities and their relationships]. [Typically, more than two nodes are used]. \href{https://www.davidrumsey.com/luna/servlet/detail/RUMSEY~8~1~28083~1120232}{Example: ``General questions''}.                                                                                                                                                \\
        \midrule
        \leafTaxon{Sankey diagram}                 & A [visual representation] that uses [bands with varying width] to represent [quantities of flows between entities in a network]. [The width of each band is proportional to the quantity]. \href{https://www.davidrumsey.com/luna/servlet/detail/RUMSEY~8~1~309268~90079194}{Example: ``Carte Figurative du Tonnage des Chemins de Fer Francais en 1884''}.                                                                                     \\
        \midrule
        \leafTaxon{timeline}                       & A [visual representation] that commonly uses [points, intervals, and textual labels arranged along a time axis] to represent [chronological sequences of events or durations over time]. \href{https://www.davidrumsey.com/luna/servlet/detail/RUMSEY~8~1~29681~1140683}{Example: ``Table, souverains de l'Europe''}.                                                                                                                           \\
        \midrule
        \leafTaxon{wind rose}                      & A [visual representation] that uses [radially arranged spokes with varying lengths] to represent [the speed and direction of wind]. [Each spoke corresponds to a compass direction]. \href{https://www.davidrumsey.com/luna/servlet/detail/RUMSEY~8~1~346720~90114207}{Example: ``Pilot chart of the North Atlantic Ocean. March 1885''}.                                                                                                       \\
        \midrule
        \leafTaxon{other statistical chart}        & A [visual representation] that [represents statistical data and does not fall into other categories]. \href{https://www.davidrumsey.com/luna/servlet/detail/RUMSEY~8~1~263594~5524292}{Example: ``Comparative chart, of Continents, Oceans, Islands \&c''}.                                                                                                                                                                                     \\
        \midrule
        \leafTaxon{rose chart}                     & A [visual representation] that uses [radially arranged wedges with varying radii] to represent [quantitative data distributed across categories]. \href{https://www.davidrumsey.com/luna/servlet/detail/RUMSEY~8~1~309778~90078927}{Example: ``Carte Figurative des Donnees Relatives A L'Entrentien des Routes Nationales en 1878''}.                                                                                                          \\
        \midrule
        \leafTaxon{parallel coordinates}           & A [visual representation] that uses [lines connecting points across parallel axes] to represent [multi-dimensional data]. [Each axis corresponds to a dimension, and each line represents a data point]. \href{https://archive.org/details/b31366661/page/n112}{Example: ``Statistical Atlas of the United States 1900 - Plate No. 21''}.                                                                                                       \\
        \midrule
        \leafTaxon{radar chart}                    & A [visual representation] that uses [lines connecting points across radially arranged axes] to represent [multi-dimensional data]. [Each axis corresponds to a dimension, and each line represents a data point]. \href{https://www.davidrumsey.com/luna/servlet/detail/RUMSEY~8~1~32168~1151510}{Example: ``Deaths scarlet fever, diphtheria, croup 1900, 1890''}.                                                                             \\
        \midrule
        \leafTaxon{treemap}                        & A [visual representation] that uses [nested rectangles with varying sizes] to represent [a hierarchical structure with associated quantitative data]. [The area of each leaf rectangle is proportional to its quantitative value within the hierarchy]. \href{https://archive.org/details/CAT10508326/page/n11}{Example: ``Forest service atlas. Extracts from the statistical volume of the Forest Atlas for the fiscal year 1907 - Page 1''}. \\
        \midrule
        \leafTaxon{violin plot}                    & A [visual representation] that commonly uses [areas symmetrical to a central axis] to represent [the distribution of a continuous variable]. \href{https://www.davidrumsey.com/luna/servlet/detail/RUMSEY~8~1~20763~560044}{Example: ``U.S. population by altitude, 1880, 1890''}.                                                                                                                                                              \\
        \bottomrule
    \end{tabular}
\end{table*}

    \ifx\hidemain\undefined
    \else
        \bibliographystyle{abbrv-doi-hyperref}
        \bibliography{assets/bibs/papers,assets/bibs/historical-visualizations}
    \fi
\fi

\end{document}